\documentclass[useAMS,usenatbib,usegraphicx]{mn2e}
\newcommand{\HI}{H{\sc~i}{ }}
\newcommand{\HII}{H{\sc~ii}{ }}

\title[Impact of UV radiation on dwarf galaxies]{Impact of UV radiation from giant spirals on the evolution of dwarf galaxies}
 
\author[S. Mashchenko, C. Carignan and A. Bouchard]{S. Mashchenko,$^{1,2}$ C. Carignan,$^2$ 
and A. Bouchard$^{2,3}$\\
$^1$Department of Physics and Astronomy, McMaster University, Hamilton, ON, L8S 4M1, Canada\\
$^2$D\'epartement de physique and Observatoire du mont M\'egantic, 
Universit\'e de Montr\'eal, C.P. 6128, Succ. Centre-ville, Montr\'eal,\\
 Qu\'ebec, H3C 3J7, Canada\\
$^3$Australia Telescope National Facility, PO Box 76, Epping, NSW 1710, Australia
}

\begin{document}

\maketitle

\begin{abstract}
We show that ultraviolet radiation, with wavelengths shorter than 2000~\AA,
escaping from the disks of giant spirals could be one of the principal factors
affecting the evolution of low mass satellite galaxies. We demonstrate, using an
analytical approach, that the Lyman continuum part of the radiation field can
lead to the ionization of the ISM of dwarf galaxies through the process of
photoevaporation, making the ISM virtually unobservable. The FUV part
($912<\lambda<2000$~\AA) is shown to dominate over the internal sources of
radiation for most of the Galactic dwarf spheroidals.  The proposed
environmental factor could be at least partially responsible for the bifurcation
of the low mass proto-galaxies into two sequences --- dwarf irregulars and dwarf
spheroidals. We discuss many peculiarities of the Local Group early-type dwarfs
which can be accounted for by the impact of the UV radiation from the host
spiral galaxy (Milky Way or M31).
\end{abstract}


\section{INTRODUCTION}

The diverse group of low luminosity (M$_{\rm V}>-12\fm 5$) Local Group dwarf
galaxies is loosely divided (based on their recent star formation history, SFH,
and neutral gas content) into three categories: dwarf irregulars (dIrr),
intermediate type dwarfs (dIrr/dSph), and dwarf spheroidals (dSph). Despite some
obvious differences, these galaxies do share a few important properties: (a)
they are pressure supported (rotation is dynamically unimportant); (b) for a
given luminosity, they have comparable spatial extent (especially when
considering the distribution of the old stars); (c) they have comparable low
metallicity $[{\rm Fe/H}]\sim -2$; (d) in most cases, they have very aspheric
shapes with ellipticity $e \sim 0.2-0.4$; (e) the available kinematic data
suggest that they are dark matter (DM) dominated.

Local Group dSph galaxies are known to have a wide range of SFHs --- from being
consistent with a single burst scenario (e.g. Ursa Minor) to the very complex
multiple bursts case of Carina \citep{gre97}. Low luminosity dSphs have not
formed stars for at least a Gyr. Intermediate type (dIrr/dSph) Local Group
dwarfs (LGS~3, Antlia, and Phoenix) had their most recent star bursts $\sim
100$~Myr ago \citep{mil01,pie99,mga99}. Intrinsically faint dIrr galaxies are
all forming stars at the present time but with low efficiency (${\rm d}m/{\rm
d}t < 0.001$~M$_{\odot}$~yr$^{-1}$, \citealt{mat98}).

A few scenarios have been proposed to explain the complex SFH of dSph galaxies,
including episodic accretion of intergalactic gas \citep*{sil87}, ISM heating by
SN~Ia leading to prolonged periods of time with no star formation \citep{bur97},
and bar induced star bursts in the tidally stirred dwarfs scenario of
\citet{may01}.

Aside from the differences in the present day star formation rate (SFR), low
luminosity galaxies from the sequence dIrr -- dIrr/dSph -- dSph differ in their
neutral gas content. Local Group dIrrs are gas rich: the ratio of the \HI mass
to the V-band luminosity $M_{\rm HI}/L_V$ ranges from $\sim 1.4$ for GR~8 to
$\sim 2.8$ for DDO~210 \citep{mat98}. Intermediate type dwarfs have smaller \HI
content: from $M_{\rm HI}/L_V\sim 0.14$ for Phoenix \citep{stg99,gal01} to
$M_{\rm HI}/{\rm L}_{\rm V}\sim 0.4$ for LGS~3 \citep{mat98} and Antlia
\citep{bar01}. Dwarf spheroidals have no or little H{\sc~i}. The only possible detection with
the \HI emission located within the optical extent of the dSph and the radial
velocity of the gas being within $15$~km~s$^{-1}$ from the optical velocity of
the dwarf is that of Sculptor ($M_{\rm HI}/L_V\sim 0.09$,
\citealt*{car98,bcm03}).

The location of the intrinsically faint Local Group dwarfs on the dSph --
dIrr/dSph -- dIrr sequence appears to correlate with their proximity to large
spirals \citep[e.g.][]{ber99}: dSphs are concentrated in the vicinity of the
Milky Way and M31, low mass dIrrs are isolated systems, and the intermediate
type dwarfs are somewhere in between. Environment appears to bear a significant
impact on the dwarfs evolution.

We propose a novel mechanism which can explain the differences in SFHs and
neutral gas content between dSphs and dIrrs. Our hypothesis is that the
electromagnetic radiation (especially Lyman continuum, LyC, and far ultraviolet,
FUV) escaping from the host spiral galaxy can play a decisive role in the
evolution of dSphs. Significant amounts of gas can be ionized and heated by the
host galaxy LyC when dSphs are on the relatively high galactic latitude parts of
their orbits. FUV can also play a role by preventing the formation of the cold
neutral medium (CNM) phase of the ISM, which presence is believed to be required
for star formation.  Moving along their orbits around the host galaxy, dSphs
spend relatively short periods of time near the plane of the host, where the
fluxes of the LyC and FUV radiation drop to lower metagalactic levels.  For some
of the dwarfs, the time spent in the shadow produced by the \HI disk of the host
spiral is enough for their ISM to recombine, cool down and form stars in a short
burst. The timescale for the repeated star bursts in this scenario is equal to
half of the orbital period, or $\sim 1-5$~Gyr, which is in agreement with the
observed SFHs. dSph galaxies with total masses $<2\times 10^8$~M$_{\odot}$
gradually lose their ISM when the gas is ionized and heated to $\sim 10^4$~K by
the LyC radiation from the host galaxy, leading to the globally declining SFR.
This should affect more the dwarfs on almost polar orbits, because their ISM is
being kept ionized for the largest fraction of their lifetimes.

This paper is organized as follows. Section~\ref{Theory} presents our analytical
model of the photoionization of the ISM of dSphs by the LyC radiation from the
host spiral galaxy, and evaluates the importance of the external FUV radiation
for Galactic dSphs. Section~\ref{Add} gives the observational evidence for the
impact of the electromagnetic radiation from giant spirals on the evolution of
dwarf satellite galaxies.  Section~\ref{Discussion} discusses the
implications of our results and gives our conclusions.

\section{MODEL}
\label{Theory}

\subsection{LyC radiation: a toy model}
\label{LyC}

We adopt the following very simple description of the Galactic ionizing
radiation field. Total LyC luminosity of the Galaxy is $L_0=3\times
10^{53}$~s$^{-1}$. The escape fraction of LyC photons is $f_{\rm esc}=0.1$ in
all directions except for a narrow zone of $\pm 15\degr$ near the Galactic plane
where it is equal to zero. The background LyC flux is $F_{bg}=
10^4$~cm$^{-2}$~s$^{-1}$.  In this model the angle-averaged value of the escape
fraction is $\langle f_{\rm esc}\rangle\simeq 0.07$, which is consistent with
the observational upper limits on the metagalactic ionizing background
\citep*{shu99,bia01}.  The derived radius of the sphere centred on the Milky Way
where the Galactic ionizing flux dominates the metagalactic flux is 224~kpc.

First we consider the following toy model for the ISM of a dSph: a homogeneous
isothermal \HI sphere with the initial radius $R_0=500$~pc, temperature
$T_0=(5-10)\times 10^3$~K, and number density $n_0=0.01$~cm$^{-3}$.  The gas
pressure is $P_0/k=50-100$~K~cm$^{-3}$.  The temperatures of $(5-10)\times
10^3$~K are typical for a warm neutral medium (WNM) phase of the ISM
\citep{wol95}.  The mass of the gas is $M_{\rm HI}=1.3\times
10^5$~M$_{\odot}$.  The \HI column density through the centre of the
cloud is $N_0=3.1\times 10^{19}$~cm$^{-2}$.

For Galactic satellites with measured proper motions (LMC, Sculptor, Ursa Minor,
and Fornax), the radial component of the space velocity in the Galactic frame of
reference is much smaller than the tangential component \citep*{sch97,pia02},
suggesting low eccentricity orbits.  Integration of orbits of the satellites in
the Galactic isothermal halo potential gives a small ratio of apocentric to
pericentric distances $R_a/R_p\simeq 2-2.4$ \citep{joh98}. For simplicity,
for the rest of the paper we will assume that the orbits of the Galactic
satellites have zero eccentricity.

We place our toy dSph on a circular orbit around the Milky Way at a distance of
$R=100$~kpc, which is the distance of Carina --- the dwarf spheroidal with
probably the most puzzling star formation history consisting of three star
bursts separated by 4-8~Gyr \citep*{hur98}.  In our model, the Galactic ionizing
flux at such a distance is $2.5\times 10^4$~cm$^{-2}$~s$^{-1}$. When we include
the background radiation, the total LyC flux rises to $F_{tot}=3.5\times
10^4$~cm$^{-2}$~s$^{-1}$.

Can such a low LyC flux photoionize \HI in our toy model? The answer appears to
be ``no'' as the flux of LyC photons $F_{tot}$ is less than the recombination
rate $F_0$ along a line crossing the centre of the cloud:
$F_0=2R_0n_0^2\alpha_0^{(2)}=8\times 10^4$~cm$^{-2}$~s$^{-1}$. (Here
$\alpha_0^{(2)}\simeq 2.6\times 10^{-13}$~cm$^3$~s$^{-1}$ at $10^4$~K is the
coefficient of recombination to all but the ground level of the hydrogen atom.)

\begin{figure}
\includegraphics[scale=0.45]{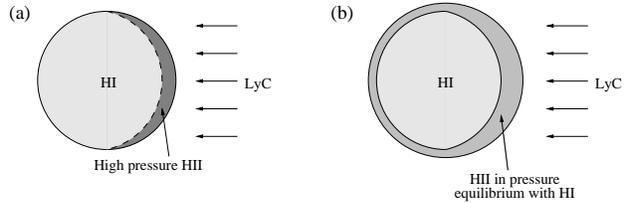}
\caption{Photoevaporation of the ISM of a dSph galaxy exposed
to the LyC photons escaping from the disk of the host galaxy: (a) on the side
facing the source of the ionizing radiation a thin layer of the high pressure
photoionized hydrogen is formed; (b) trying to reach a pressure equilibrium with
the neutral gas, the \HII gas expands and fills the outer regions of the galaxy,
allowing a fresh layer of \HI to be ionized. The process repeats itself until
the \HI gas is fully ionized, or until the ionized gas on the irradiated side of
the cloud becomes optically thick to the LyC photons.
\label{fig_Evap}}
\end{figure}

An important caveat is that in an \HI cloud exposed to anisotropic
ionizing radiation field the gas which is being photoionized does not stay
at its original location.  The increased pressure pushes the
photoionized gas away, exposing new \HI to the incident LyC
radiation. This process termed photoevaporation has been studied both
analytically \citep{ber89,ber90} and numerically
\citep{lef94} in application to an
interstellar cloud exposed to the ionizing flux from a nearby hot star.
This theory is not directly applicable to the
case of the ISM of a dSph galaxy because of the invalidity of the following
assumptions adopted by the authors: 1) the temperature of the neutral gas is
much lower than the temperature of the ionized gas $T_{\rm HII}\simeq 10^4$~K;
2) initially the cloud is in pressure equilibrium with the intercloud medium; 3)
the ionized gas is not gravitationally bound to the cloud, and can freely flow
away.

Nevertheless, we argue that the process of photoevaporation can also take place
in the ISM of a dwarf galaxy. Even in the case of the highest possible WNM
temperature $T_{\rm HI}\simeq 10^4$~K, the pressure of the photoionized gas is
$\sim 2$ times higher than the original pressure of \HI (because during the
photoionization of hydrogen the number of particles doubles, and the temperature
is maintained at the $T_{\rm HII}\simeq 10^4$~K level when the heating by the
LyC photons is balanced by radiative cooling). Unlike the interstellar cloud
case, in photoevaporating ISM of a reasonably massive dwarf galaxy the ionized
gas can stay gravitationally bound to the galaxy. Higher pressure causes the
\HII gas to expand, filling the outer regions of the galaxy, allowing a fresh
layer of \HI to be ionized (Fig.~\ref{fig_Evap}). The speed with which the
ionization front propagates inside the cloud is determined by how fast the
photoionized gas can flow away and be redistributed in the outer parts of the
dwarf galaxy, and is expected to be less than or comparable to the isothermal
sound speed in the \HII region $c_g=11.7$~km~s$^{-1}$.

Under the assumption that during photoionization the \HII gas reaches a state of
pressure equilibrium with the \HI gas (so the final pressure $P_1$ is equal to
the initial pressure $P_0$), the parameters for the fully photoionized cloud are
$R_1=\chi^{1/3}\,R_0$, $n_1=\chi^{-1}\,n_0$, $N_1=\chi^{-2/3}\,N_0$, and
$F_1=\chi^{-5/3}\,F_0$. (Here $\chi\equiv 2\,T_1/T_0$.) Assuming the
\HII temperature of $T_1=10^4$~K, we obtain $\chi=2-4$, so the final radius $R_1$,
number density $n_1$, column density $N_1$ and minimum LyC flux $F_1$ required
to keep the cloud fully ionized are $R_1=630-790$~pc, $n_1=(2.5-5)\times
10^{-3}$~cm$^{-3}$, $N_1=(1.2-1.9)\times 10^{19}$~cm$^{-2}$, and
$F_1=(0.8-2.5)\times 10^4$~cm$^{-2}$~s$^{-1}$.  The photoevaporation time scale
is of the order of $2R_0/c_g=84$~Myr. Once the dwarf moves in the shadow produced by
the Galactic \HI layer, the gas will recombine with a time scale of $\tau_{\rm
rec}=1/(\alpha_0^{(2)}n_1)=24-49$~Myr.  Both photoevaporation and recombination
time scales are a small fraction of the dSph's orbital periods of a few Gyr.

Assuming that the process of photoevaporation is 100\% efficient, in our toy
model the ISM of a dwarf located $\sim 100$~kpc away from the Milky Way can be
fully photoionized by the Galactic LyC photons: $F_1<F_{tot}$. Once the dwarf
moves into the shadow produced by the Galactic disk, the gas can recombine on a
relatively short time scale, potentially leading to an episode of star
formation.

\subsection{LyC radiation: more realistic approach}

A homogeneous gas sphere is admittedly not a very realistic model for the ISM of
dSphs. In our next step we consider the distribution of isothermal ionized
hydrogen gas being in hydrostatic equilibrium inside properly normalized DM
halos with different masses and density profiles (either NFW, or
\citealt{bur95}).  The temperature of the gas is
$T_1=10^4$~K, so the sound speed is $c_g=11.7$~km~s$^{-1}$.  Similarly to the
toy model, we wish to find the lowest ionizing flux $F_1$ sufficient to keep the
ISM fully photoionized by estimating the recombination rate along a line
crossing the centre of the cloud.

The analytical formulae describing the hydrostatic equilibrium distribution of
isothermal gas inside NFW and Burkert's DM halos were derived by
\citet*[Appendix~A]{ste02}.  For NFW halos, these authors give the
following expressions for the dimensionless DM density $f_\rho$,
enclosed DM mass $f_M$, and gas density $f_{\rm gas}$:

\begin{eqnarray}
f_\rho&=&1/[x(1+x)^2],\\
f_M&=&3[\ln(1+x)-x/(1+x)],\label{N2b}\\
f_{\rm gas}&=&e^{-3(v_s/c_g)^2}(1+x)^{(v_s/c_g)^2/x}.
\end{eqnarray}

\noindent Here $x\equiv r/r_s$ is the dimensionless radius. The corresponding
equations for Burkert's halos are

\begin{eqnarray}
f_\rho&=&1/[(1+x)(1+x^2)],\\
f_M&=&(3/2)\{[\ln(1+x^2)]/2+\ln(1+x)-\tan^{-1}x\},\label{N3b}\\
f_{\rm gas}&=&[e^{-(1+1/x)\tan^{-1}x}\times\nonumber \\
&&\hspace{0pt}(1+x)^{(1+1/x)}(1+x^2)^{(1/x-1)/2}]^{(3/2)(v_s/c_g)^2}.
\end{eqnarray}

\noindent From equations (16), (17) and (19) of \citet{ste02} we obtained
the following approximate expressions for the scaling radius $r_s$, the scaling
velocity $v_s$, and the concentration parameter $x_{\rm vir}\equiv r_{\rm
vir}/r_s$ for DM halos in $\Lambda$CDM cosmology (assuming $\Omega_m=0.3$,
$\Omega_\Lambda=0.7$, $H_0=70$~km~s$^{-1}$~Mpc$^{-1}$, and $\sigma_8=1$):

\begin{eqnarray}
r_s&=&0.96 \left( \frac{m_{\rm vir}}{10^9 {\rm M}_\odot}\right)^{0.41} \,\, {\rm kpc},\label{eq_rs}\\
v_s&=&24.7 \left( \frac{m_{\rm vir}}{10^9 {\rm M}_\odot}\right)^{0.31}\,\, {\rm km~s}^{-1},\\
x_{\rm vir}&=&26.2\left( \frac{m_{\rm vir}}{10^9 {\rm M}_\odot}\right)^{-0.07}.\label{eq_xvir}
\end{eqnarray}

\noindent Here $r_{\rm vir}$ and $m_{\rm vir}$ are the virial radius and the 
virial mass of a dwarf, respectively. Equations (\ref{eq_rs}-\ref{eq_xvir}) are
valid for $m_{\rm vir}=10^8-10^{11}$~M$_\odot$.

The total masses of the Galactic dwarf spheroidals are not known.  Even under
the simplest ``mass follows light'' assumption, most dSphs appear to be DM
dominated \citep{mat98}. The absence of tidal features caused by the
gravitational field of the Milky Way in stellar isophots of Draco down to a very
low level (0.001 of the central surface brightness) suggests that at least in
this dSph the DM halo is more extended than the stellar body
\citep{ode01}. Comparison of the structure and kinematics of Milky Way
satellites with cosmological N-body $\Lambda$CDM simulations suggests that the
dSphs may represent the most massive substructures in the Galactic DM halo, with
masses up to a few $10^9$~M$_\odot$ \citep{sto02,hay03}. By allowing stars and
DM to have different velocity distributions, \citeauthor{lok02}'
(\citeyear{lok02}) analysis of the observed radial profiles of stellar
velocity dispersion for Fornax and Draco yielded a range of possible total
masses of $\sim (1-4)\times 10^9$~M$_\odot$ for each of these dwarfs. (The
analysis included both NFW-type and flat-core-type DM profiles.)

In this paper we explore the range of virial masses of dwarf spheroidals between
$2.5\times 10^8$ and $4\times 10^9$~M$_\odot$, with a fiducial value of
$10^9$~M$_\odot$. Galaxies less massive than $\sim 2\times 10^8$~M$_\odot$ will
not be able to keep the photoionized gas gravitationally bound for a long period
of time (unless they are confined by non-negligible pressure of the hot Galactic
corona), and more massive than a few $10^9$~M$_\odot$ will be inconsistent with
the predictions of $\Lambda$CDM models.

The halos of the dSphs are truncated by the Galactic tidal field. We estimate
the tidal radius $r_t$ of satellites by solving numerically the following
non-linear equation applicable to dwarfs on circular orbits around the host
\citep{hay03}:

\begin{equation}
\label{N1}
\frac{m(r_t)}{r_t^3} = \frac{M(R)}{R^3}\left[2-\frac{R}{M(R)}\frac{\partial M}{\partial R}\right].
\end{equation}

\noindent Here $m(r)=f_M(r/r_s)v_s^2r_s/G$ and $M(R)$ are the enclosed DM mass for
the satellite and host, and $R$ is the distance of the satellite from the centre
of the host. We assume that the Milky Way DM halo is an isothermal sphere, with
$M(R)=2\sigma^2R/G$. We set $\sigma=113.6$~km~s$^{-1}$ so that $M(250~{\rm
kpc})=1.5\times10^{12}$~M$_\odot$. For the range of galactocentric distances
of dSphs ($R=70-250$~kpc), the mass of a tidally truncated satellite $m_t$ is
found to be more than half of the original virial mass $m_{\rm vir}$.

Our algorithm is as follows. For a given satellite's halo model (Burkert or NFW),
virial mass $m_{\rm vir}$, and galactocentric distance $R$, we first find the
tidal radius $r_t$ by solving the non-linear equation~(\ref{N1}). The
recombination rate $F_1$ along a line crossing the centre of the halo is given
by

\begin{equation}
\label{N2}
F_1=2\alpha_0^{(2)}n_0^2\,r_s\!\int\limits_0^{x_t}\!\! f_{\rm gas}^2(x)\,{\rm d}x.
\end{equation}

\noindent Here $x_t\equiv r_t/r_s$ is the dimensionless tidal radius and $n_0$
is the central proton number density of the gas. 

Substituting $F_1$ in equation~(\ref{N2}) with the total LyC flux at the
distance $R$ from the Galactic centre $F_{\rm tot}=F_{\rm bg}+f_{\rm
esc}L_0/(4\pi R^2)$ and solving the resulting equation for $n_0$ gives us an
estimate of the largest possible central number density $n_{0,\rm tot}$ for the
ISM being in a fully photoionized state when the satellite is located high above
the Galactic plane:

\begin{equation}
\label{N3}
n_{0,\rm tot}=\left[\frac{F_{\rm bg}+f_{\rm esc}L_0/(4\pi
R^2)}{2\alpha_0^{(2)}r_s\int\limits_0^{x_t} f_{\rm gas}^2(x)\,{\rm
d}x}\right]^\frac12.
\end{equation}

\noindent Similarly, substituting $F_1$ with $F_{\rm bg}$
will result in the estimate of the largest possible central number
density $n_{0,\rm bg}$ for the ISM being in a fully photoionized state
when the dwarf is located near the Galactic plane (in the shadow
produced by the \HI disk of the Milky Way):

\begin{equation}
\label{N4}
n_{0,\rm bg}=\left[\frac{F_{\rm bg}}{2\alpha_0^{(2)}r_s\int\limits_0^{x_t}
f_{\rm gas}^2(x)\,{\rm d}x}\right]^\frac12.
\end{equation}

\noindent The corresponding maximum total masses $m_{\rm gas,tot}$ and $m_{\rm
gas,bg}$ of the fully photoionized ISM are

\begin{equation}
\label{N5}
m_{\rm gas,tot/bg}=4\pi n_{0,\rm tot/bg}\,m_{\rm HI}r_s^3\!\int\limits_0^{x_t}\!\! f_{\rm gas}(x)\,x^2\,{\rm d}x.
\end{equation}

\noindent Here $m_{\rm HI}$ is the mass of a hydrogen atom. From 
equations~(\ref{N3}-\ref{N5}) one can see that the following inequalities hold for
any finite $R$: $n_{0,\rm tot}>n_{0,\rm bg}$ and $m_{\rm gas,tot}>m_{\rm
gas,bg}$.

\begin{figure*}
\includegraphics[scale=0.4]{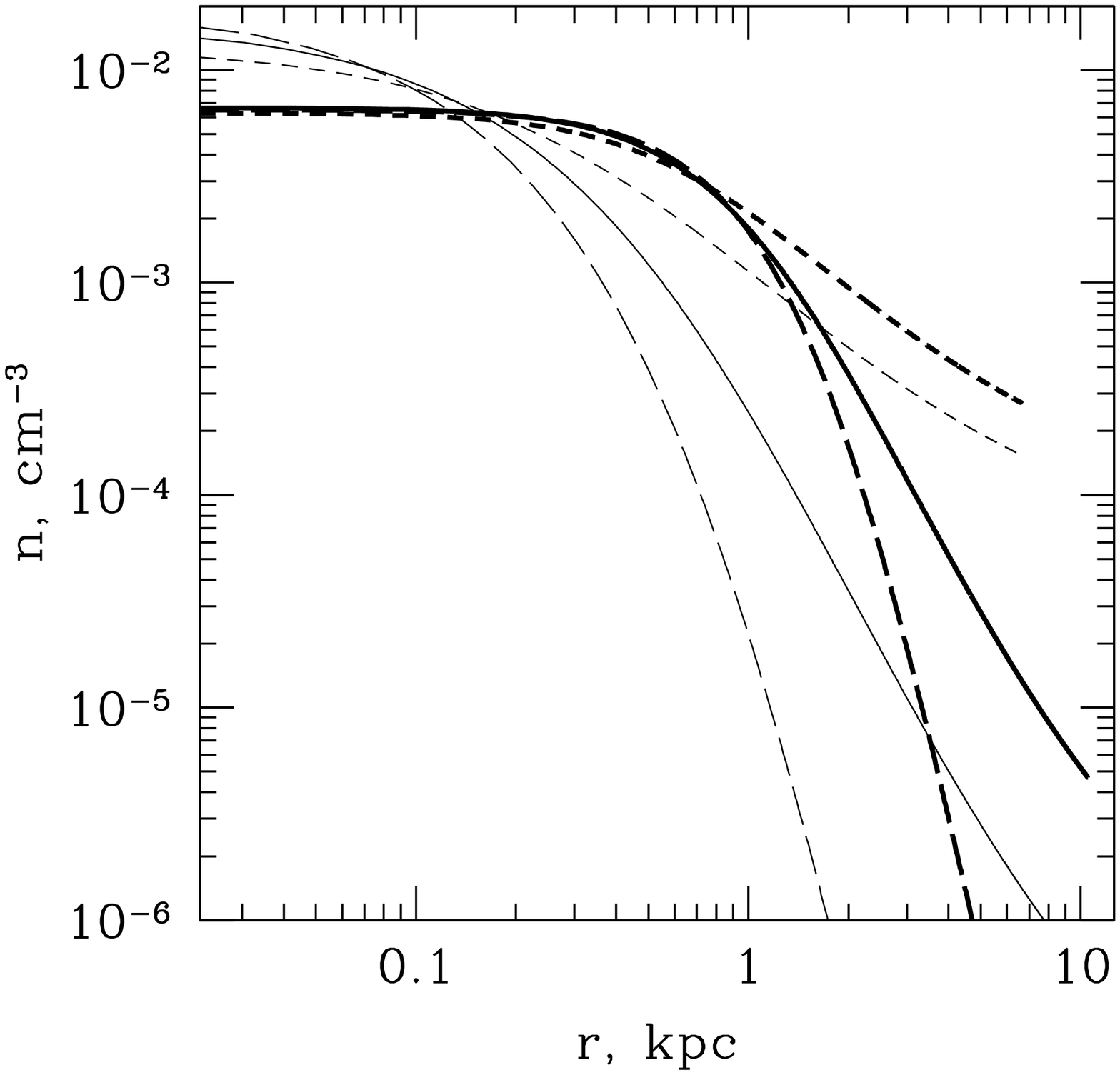}\hspace{30pt}\includegraphics[scale=0.4]{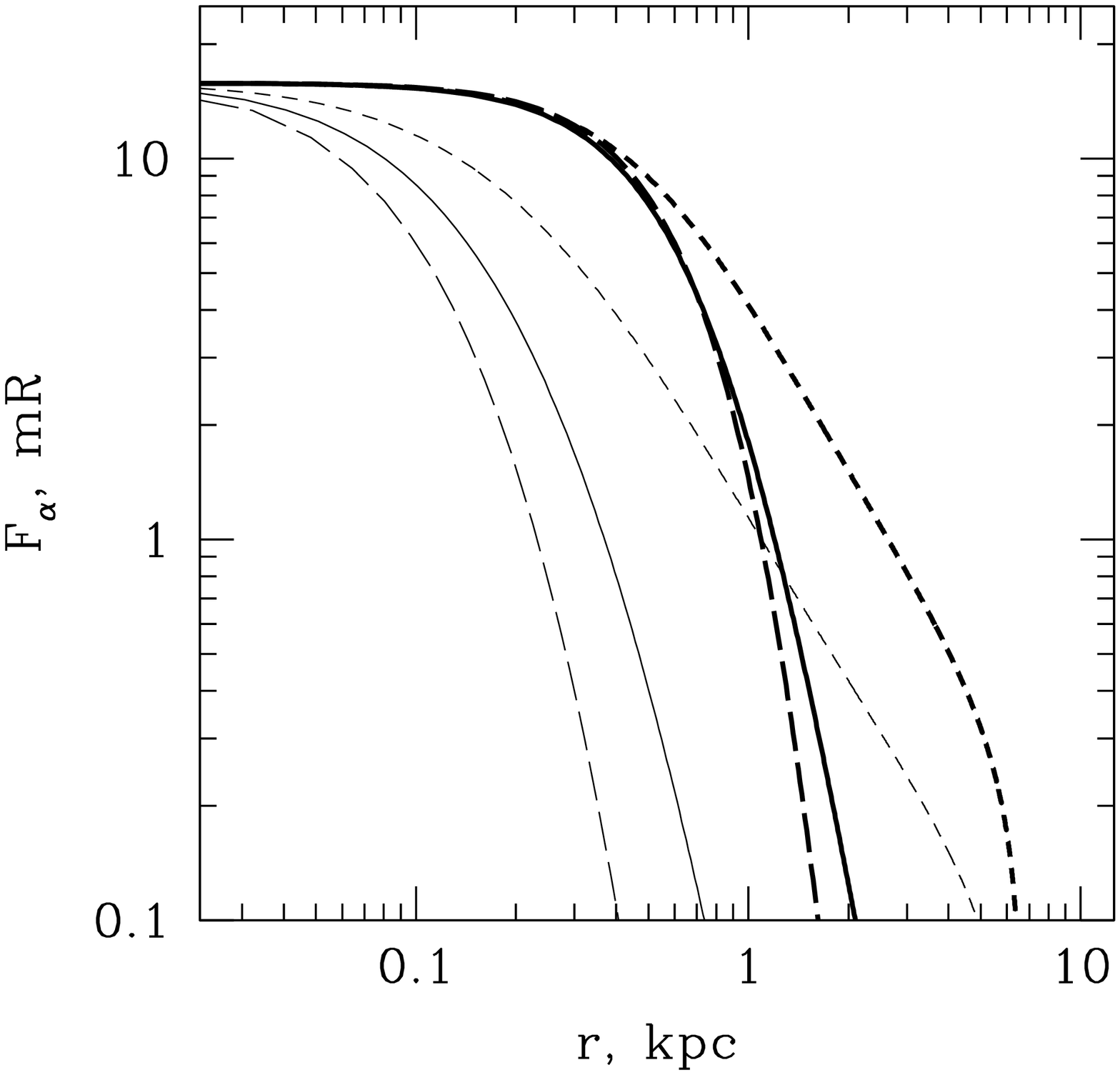}
\caption{
Radial profiles for critical ISM in Burkert (thick lines) and NFW
(thin lines) halos with virial masses of $10^9$~M$_\odot$ (solid
lines), $2.5\times 10^8$~M$_\odot$ (short-dashed lines), and $4\times
10^9$~M$_\odot$ (long-dashed lines), located 100~kpc from the Galactic
centre high above the plane. Left panel: proton number density.
Right panel: $H_\alpha$ surface brightness.
\label{ion_gas}}
\end{figure*}

In Fig.~\ref{ion_gas} we show the critical gas density profiles for different
halos located at a distance of $100$~kpc from the Milky Way (left panel). By
``critical'' we mean the ISM for which the recombination rate along a line
crossing the centre of the dwarf is equal to the incident LyC flux. The central
number density and mass for such an ISM are described by
equations~(\ref{N3}-\ref{N5}).  Subcritical (with lower $n_0$) and critical ISM
are fully photoionized, whereas a supercritical (with higher $n_0$) ISM has a
neutral core where star formation can take place. (The H$_\alpha$ surface
brightness profiles for our model halos shown in the right panel of
Fig.~\ref{ion_gas} will be discussed at the end of this section.)

As can be seen in Fig.~\ref{ion_gas}, in the lowest mass cases ($m_{\rm
vir}=2.5\times 10^8$~M$_\odot$) the photoionized gas is not well confined by the
gravity of the satellite, with the gas density at the tidal radius being only
$20-80$ times lower than the central density. On the other hand, the gas
pressure at the tidal radius is low --- around 4~K~cm$^{-3}$, so even very
tenuous Galactic corona with temperature of $10^6$~K and proton number
density of $2\times 10^{-6}$~cm$^{-3}$ would suffice to confine the gas.

In Fig.~\ref{ion_gas} one can also see that the isothermal gas density
distribution for Burkert halos has a large core with an almost constant radius
of $\sim 800$~pc for the whole range of virial masses considered. On the other
hand, the ISM in NFW halos is more centrally peaked, which is the result of their
cuspy DM density profile.

\begin{figure*}
\includegraphics[scale=0.4]{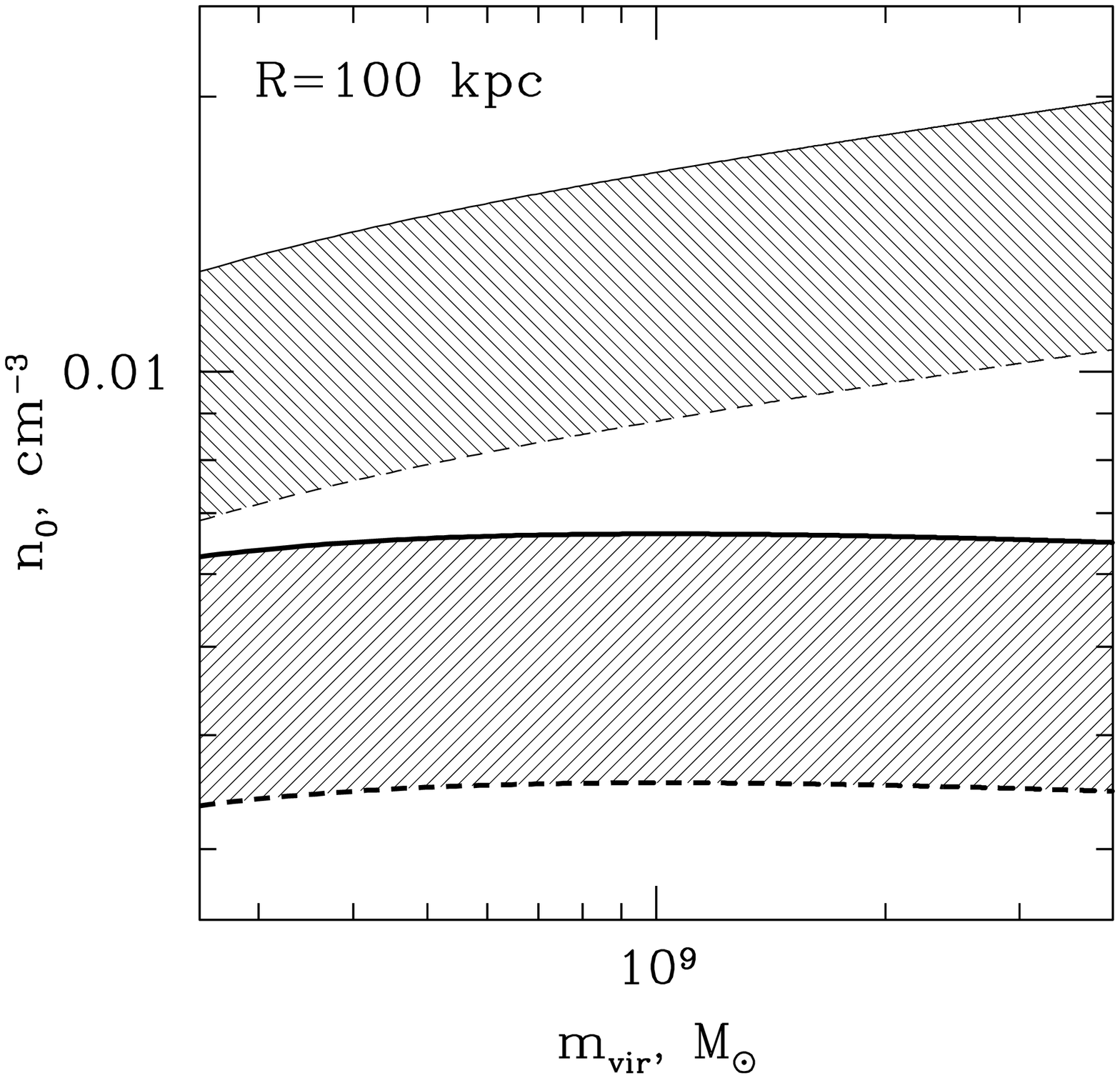}\hspace{30pt}\includegraphics[scale=0.4]{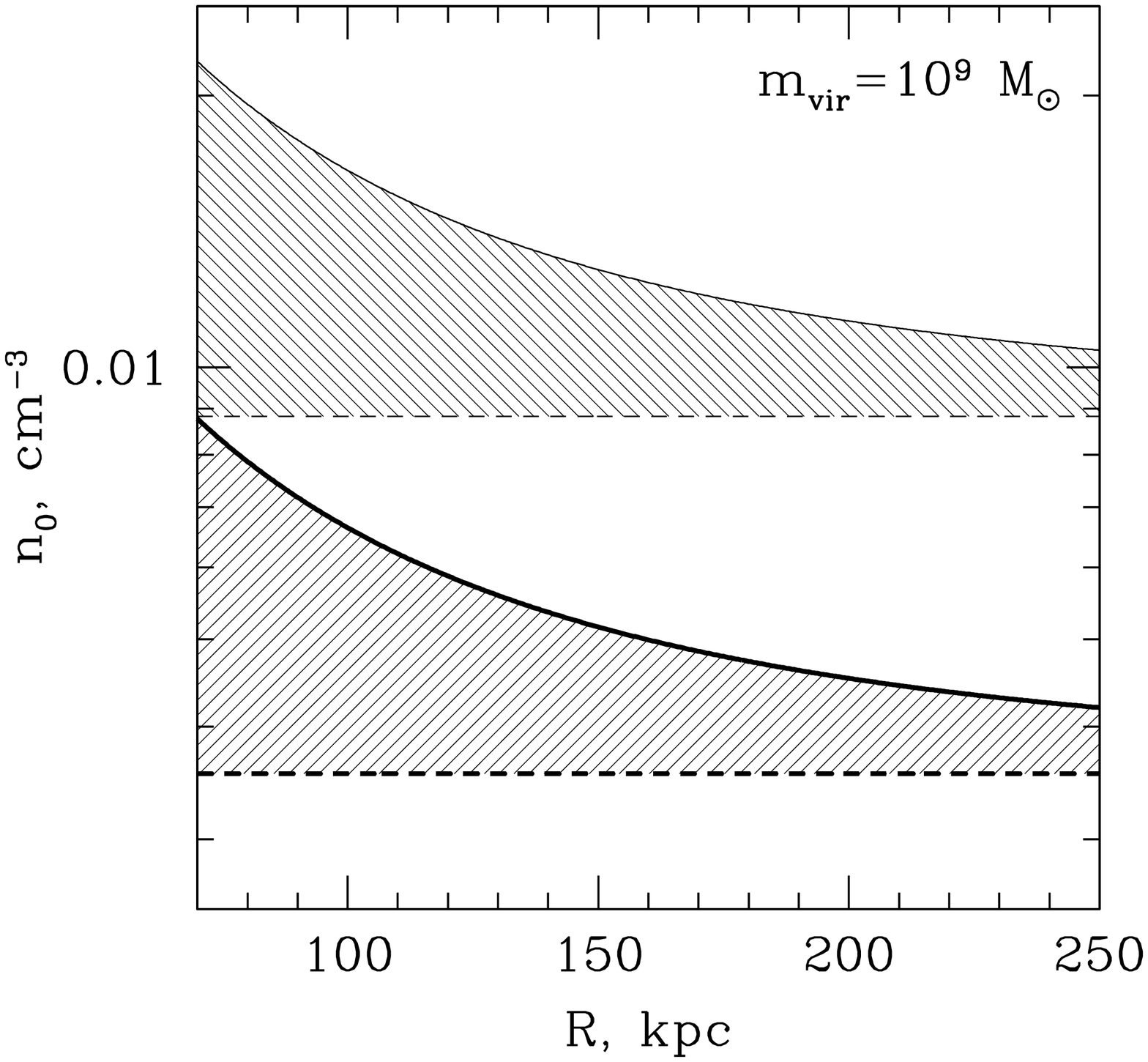}
\caption{
Critical central number density for Burkert (thick lines) and NFW (thin lines)
halos as a function of the virial mass (left panel) and galactocentric distance
(right panel). Both $n_{0,\rm tot}$ (solid lines) and $n_{0,\rm bg}$ (dashed
lines) are shown.  The shaded areas show the locus of half-critical halos.
\label{ion_n0}}
\end{figure*}

Fig.~\ref{ion_n0} shows critical central gas density for either Burkert or NFW
halos with different virial masses, located at different distances from the
Milky Way, and exposed either to the total (Galactic $+$ metagalactic) ionizing
flux, or only to the background LyC radiation (when the halo is in the shadow
produced by the Galactic disk). The shaded areas correspond to halos which are
subcritical when they are exposed to the Milky Way ionizing radiation, and
supercritical when they are in the ``LyC shadow'' and can potentially form
stars. These {\it half-critical} systems are of utmost interest to us as they
can explain both the multiple star bursts history of some of the Galactic dSphs and
the apparent absence of ISM in all of them (with the possible exception of Sculptor,
\citealt{bcm03}) including such dwarfs as Carina and Fornax which have had
some star formation in the last Gyr.

By analyzing Fig.~\ref{ion_n0} we can make a few interesting
observations. First, the range of central densities for half-critical dwarfs is
almost independent on the virial mass of the halo, especially for Burkert
halos. Second, this range is relatively large (a factor of 2 or more) for dwarfs
located $\la 100$~kpc from the Galaxy, but quickly becomes much smaller for
larger distances. At the distance of Leo I (250~kpc), the interval is so narrow,
that it would be very unlikely if the ISM of this galaxy was in the
half-critical state.  If the dwarf has any ISM, the gas is probably thin enough
to be photoionized by the metagalactic LyC background alone. In the past it
could have possessed a denser ISM with a neutral core where star formation
could have taken place.

\begin{figure*}
\includegraphics[scale=0.4]{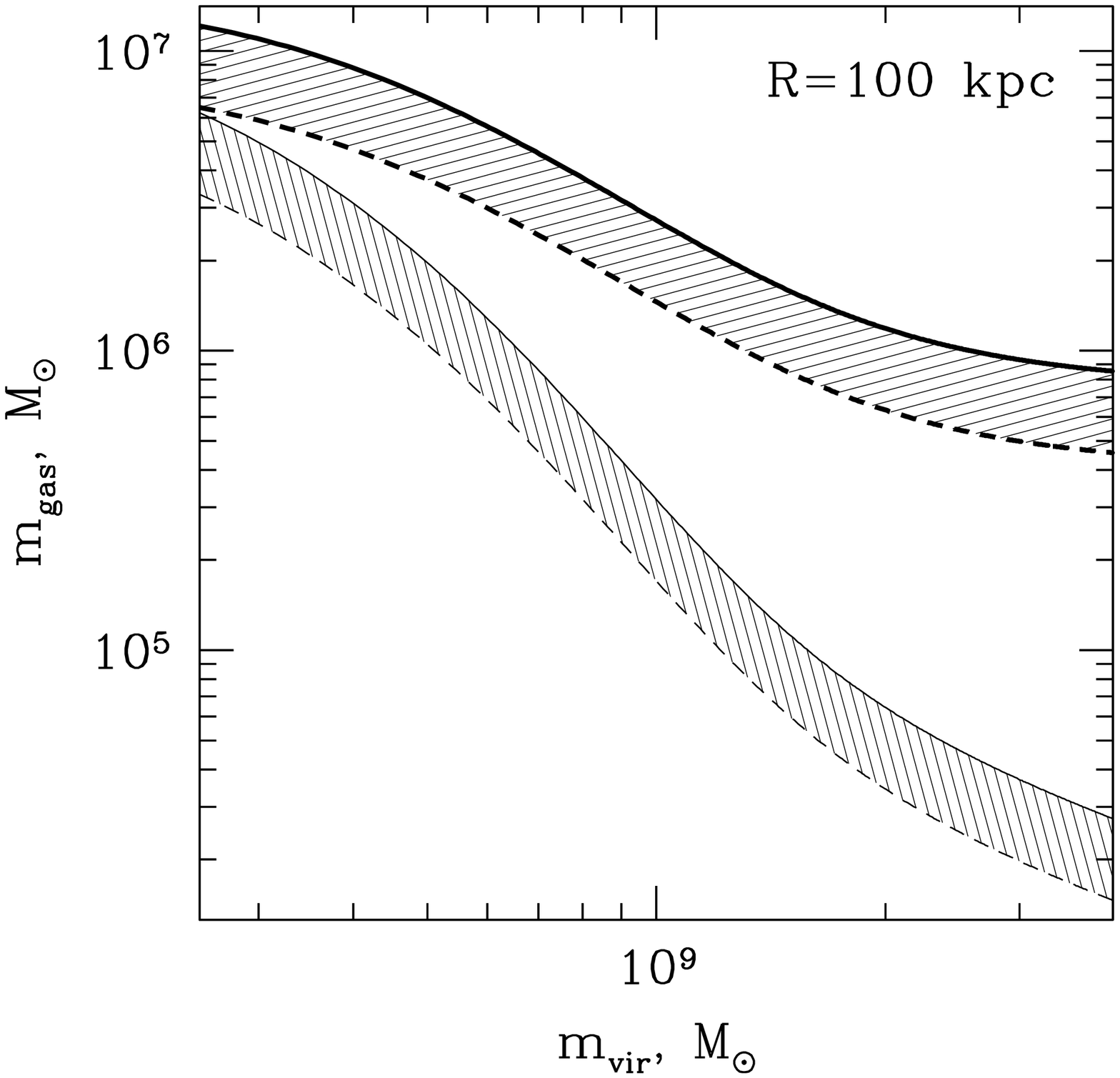}\hspace{30pt}\includegraphics[scale=0.4]{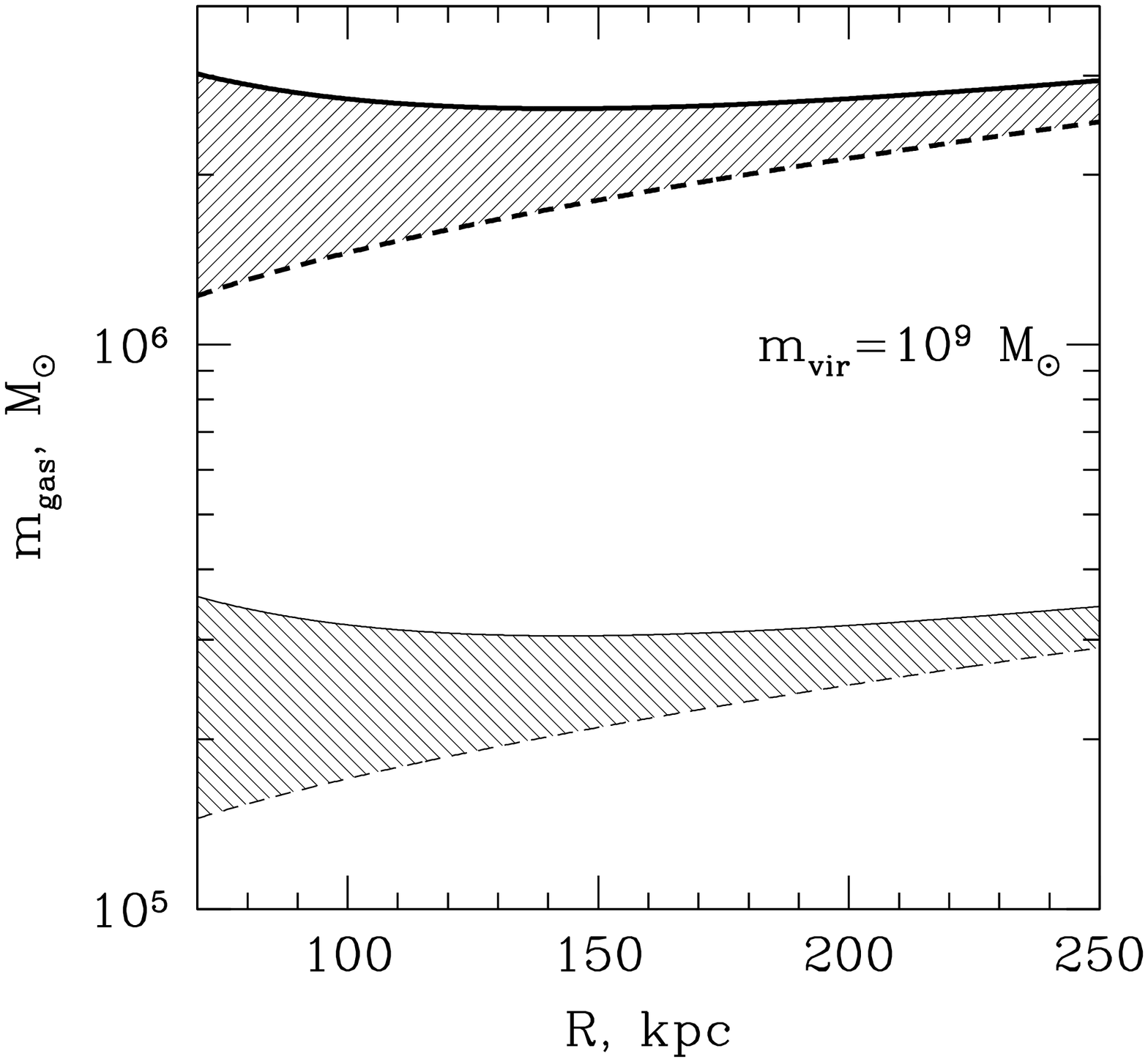}
\caption{
Critical total gas mass for Burkert (thick lines) and NFW (thin lines) halos as
a function of the virial mass (left panel) and galactocentric distance (right
panel). Both $m_{\rm gas,tot}$ (solid lines) and $m_{\rm gas,bg}$ (dashed lines)
are shown.  The shaded areas show the locus of half-critical halos.
\label{ion_mgas}}
\end{figure*}

Critical total gas masses for different halos are shown in
Fig.~\ref{ion_mgas}.  Burkert halos can contain much more massive ISM in
half-critical (shaded areas) or permanently photoionized (areas below the dashed
lines) states than NFW halos because they are less centrally concentrated. The
total mass of gas in half-critical halos is of order of $10^6$~M$_\odot$
(Burkert) or $10^5$~M$_\odot$ (NFW).  As the right panel of
Fig.~\ref{ion_mgas} suggests, isolated Burkert halos can contain millions
M$_\odot$ of gas within their virial radii kept fully photoionized by the
metagalactic ionizing radiation alone.

\begin{figure*}
\includegraphics[scale=0.4]{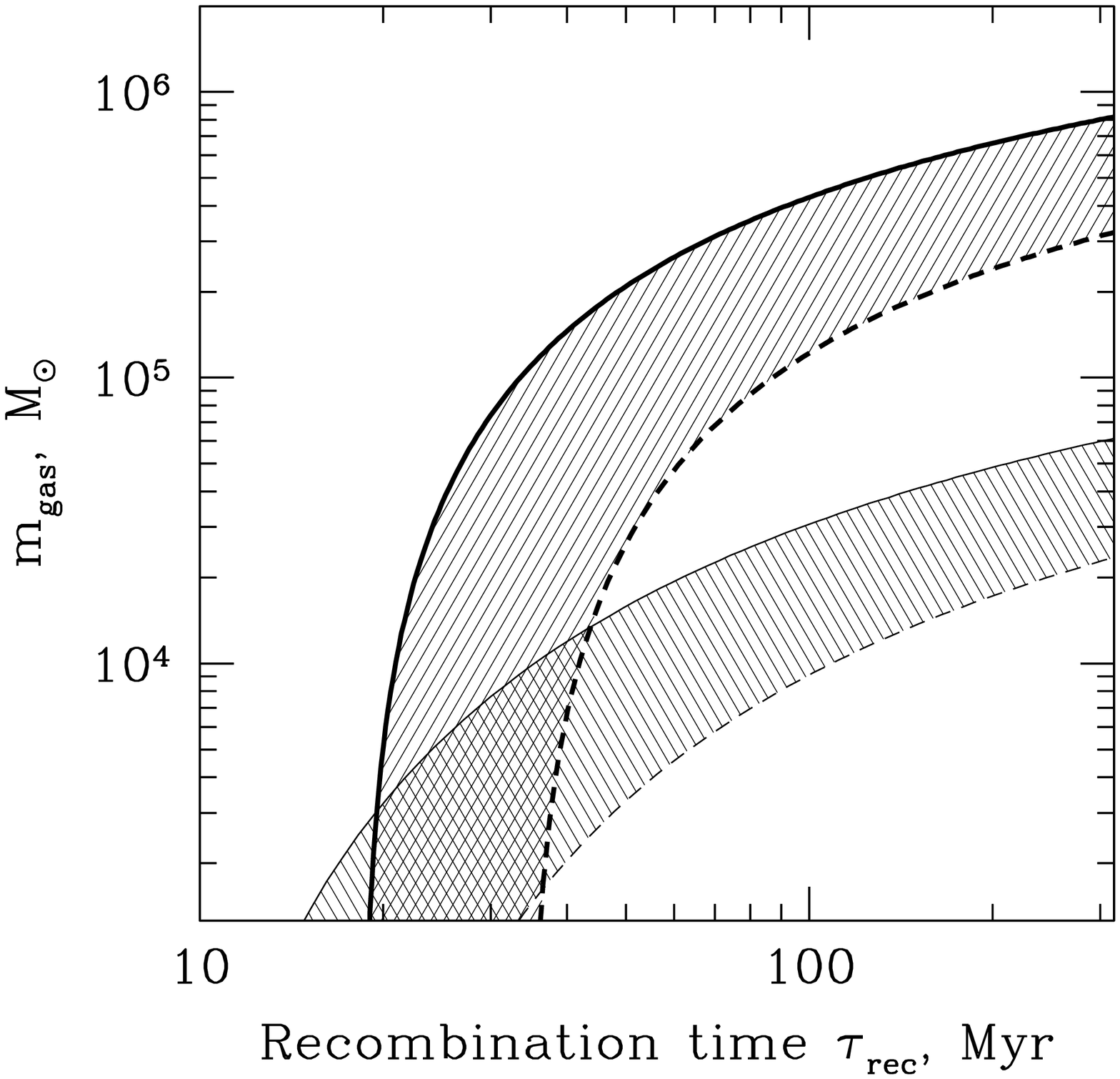}\hspace{30pt}\includegraphics[scale=0.4]{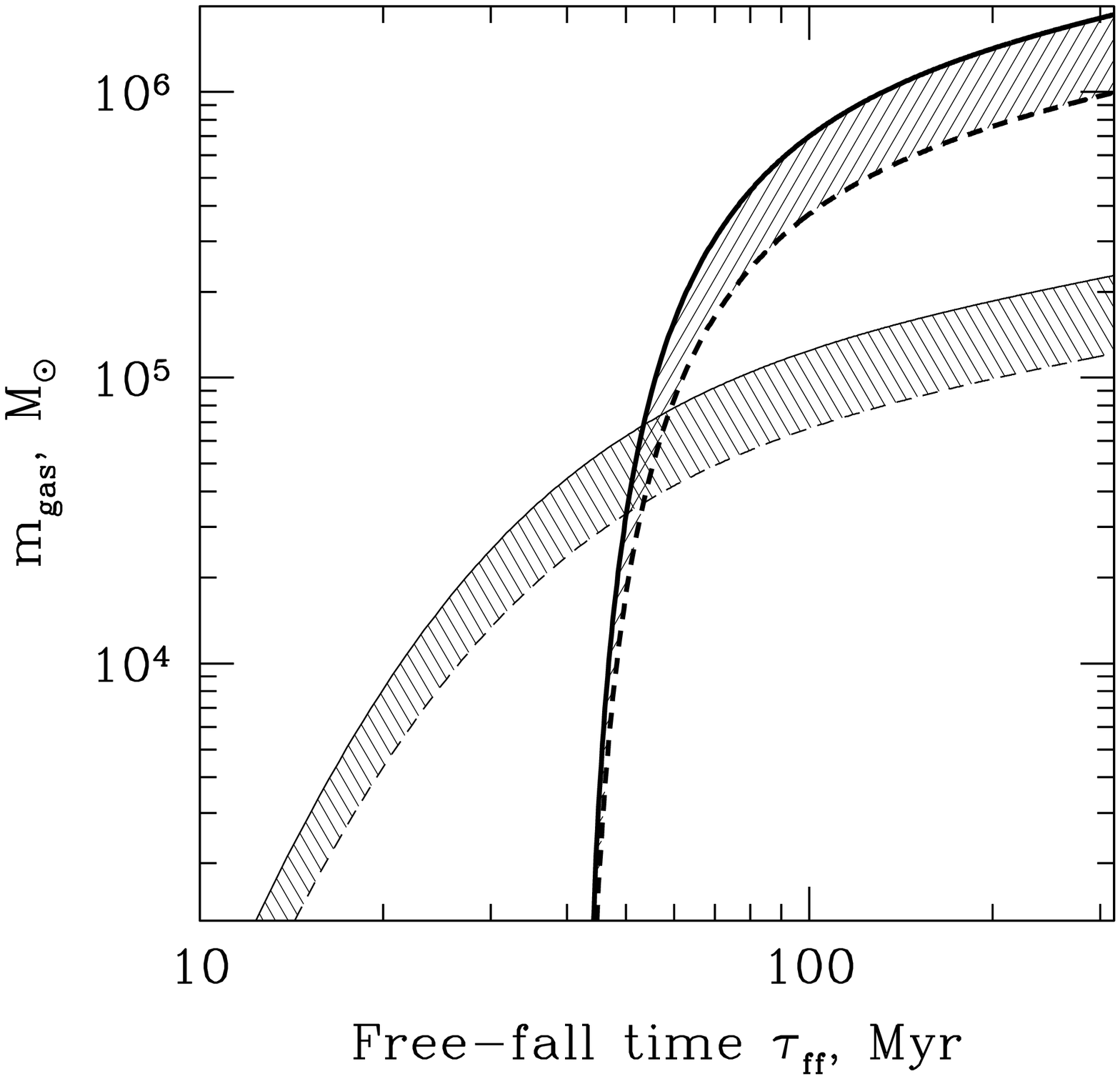}
\caption{
Mass of gas for the critical ISM in a Burkert (thick lines) or NFW (thin lines)
halo ($m_{\rm vir}=10^9$~M$_\odot$, $R=100$~kpc) which can recombine (left
panel) or collapse (right panel) within the time scale $\tau$ after the source
of the ionizing radiation is turned off. The shaded areas show the locus of
half-critical halos.
\label{ion_tau}}
\end{figure*}

Only a fraction of the total mass of gas in a half-critical halo can recombine,
cool down and collapse toward the centre of the dwarf to be available for star
formation during a relatively short passage of the satellite through the ``LyC
shadow'' cast by the Galactic disk.  In our isothermal sphere model for the
Milky Way halo, a dwarf on a circular polar orbit with $R=100$~kpc will spend
320~Myr in the shadow with the opening angle of $30^\circ$ twice per orbital
period (which is equal to 3.8~Gyr).

The shortest recombination, cooling and free-fall time scales are at the centre
of the halo.  As time goes on, gas can recombine and collapse at increasingly
larger distances from the centre of the halo. Fig.~\ref{ion_tau} quantifies
this effect for a specific case of either a Burkert or a NFW halo with a virial mass
of $10^9$~M$_\odot$ located 100~kpc away from the Galaxy.  In this figure we
show the recombination time $\tau_{\rm rec}=1/(\alpha_0^{(2)}n)$ and the
free-fall time $\tau_{\rm ff}=\sqrt{3\pi/(32G\rho)}$ \citep[p. 184]{bin94}.
(Here $\rho$ is the average total density of the system.) In our notation, the
free-fall time can be written as a function of the dimensionless radius $x$ in
the following way:

\begin{equation}
\label{N6}
\tau_{\rm ff}=\frac{\pi r_s}{v_s}\left[\frac{x^3}{8f_M(x)}\right]^{\frac12}.
\end{equation}

As can be seen in Fig.~\ref{ion_tau}, the recombination time scale is more
important than the free-fall time in controlling how much gas can become available
for star formation at the centre of the dwarf.  It is also probably more
important than the cooling time $\tau_{\rm cool}$ (not shown in
Fig.~\ref{ion_tau}). Indeed, both recombination and cooling time scales have
the same dependence on density --- $\tau_{\rm rec}=0.122/n$~Myr, and $\tau_{\rm
cool}=1.31\times 10^{-25}/(\Lambda n)$~Myr, so only for $\Lambda <1.1\times
10^{-24}$~erg~cm$^3$~s$^{-1}$ can cooling become more important than
recombination. (\citet{sut93} give the following value for the cooling function
of $10^4$~K gas: $\Lambda=4\times 10^{-24}$~erg~cm$^3$~s$^{-1}$.) In reality,
processes of recombination, cooling and collapse of the gas are not independent
from each other, and for accurate estimates of how much of cold gas can
accumulate at the centre of the halo one has to integrate a system of
appropriate differential equations, which is beyond the scope of this paper.

From Fig.~\ref{ion_tau} one can estimate that in a half-critical halo with
$m_{\rm vir}=10^9$~M$_\odot$ and $R=100$~kpc approximately $5\times
10^5$~M$_\odot$ (Burkert profile) or $5\times 10^4$~M$_\odot$ (NFW profile) of
gas can collapse to form a denser neutral core during the time spent in the
shadow ($\sim 300$~Myr). It remains to be seen if this time interval is long
enough for the formation of molecular hydrogen clouds and eventually stars. The
above mass estimates should be considered as upper limits, as the low density
gas in the outer parts of the dwarf will remain photoionized by the metagalactic
LyC radiation.

\begin{figure}
\includegraphics[scale=0.4]{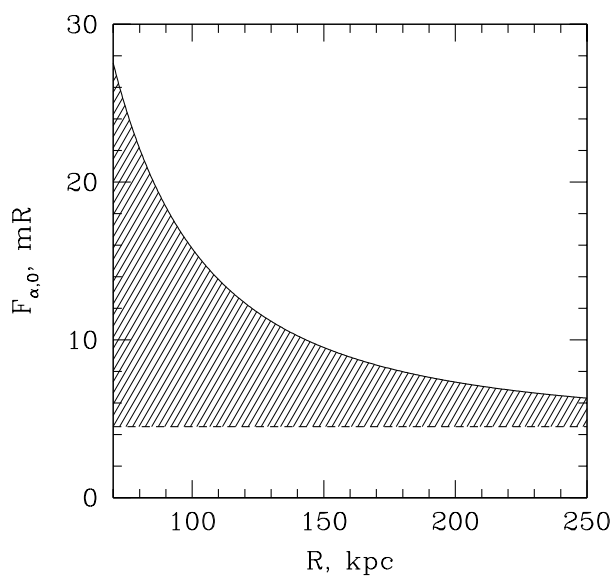}
\caption{
Range of possible central H$_\alpha$ brightness for half-critical halos located
at different distances from the Milky Way (the shaded area).
\label{ion_Ha}}
\end{figure}

The last issue we consider in this section is related to the observability of
the photoionized ISM in our model halos in the H$_\alpha$ spectral line. The
H$_\alpha$ brightness $F_\alpha$ along a line crossing the halo is related to
the corresponding recombination rate of hydrogen $F_1$ through
$F_\alpha=F_1\alpha_{32}/\alpha_0^{(2)}\simeq 0.45F_1$. Here
$\alpha_{32}=1.17\times 10^{-13}$~cm$^3$~s$^{-1}$ (for $T=10^4$~K) is the
corresponding photon production coefficient
\citep[p. 89]{spi78}.

The first observational deep upper limits on the intensity of the H$_\alpha$
emission from dwarf spheroidals set by \citet{gal03} for Draco and Ursa Minor
are 24~mR and 21~mR, respectively (1~${\rm mR}\equiv 10^3$~cm$^{-2}$~s$^{-1}$). These
results were obtained with WHAM (Wisconsin H-Alpha Mapper) which has an
effective beam size of 1$\degr$. Do these upper limits exclude the possibility of
the presence of half-critical ISM in Draco and Ursa Minor?

Fig.~\ref{ion_Ha} shows the range of central H$_\alpha$ brightness
$F_{\alpha,0}$ for half-critical halos as a function of the galactocentric
distance $R$ (the shaded area). As one can see, the expected central H$_\alpha$
flux from Galactic dSphs is very low, with values in the range of $5-30$~mR
and $5-21$~mR for the distances of Ursa Minor ($R=66$~kpc) and Draco
($R=82$~kpc), respectively.

These values of $F_{\alpha,0}$ should not be directly compared with the
observations, as the large beam size of WHAM leads to the averaging of
H$_\alpha$ flux over a significant area, effectively reducing the observed
surface brightness. This is evident from the right panel of Fig.~\ref{ion_gas}
which shows the radial profile of $F_\alpha$ for different halos. The effect
should be strongest for NFW halos, as their H$_\alpha$ flux drops significantly
on scales of $200-300$~pc ($\sim 10'$ at a distance of 100~kpc).

\begin{figure*}
\includegraphics[scale=0.4]{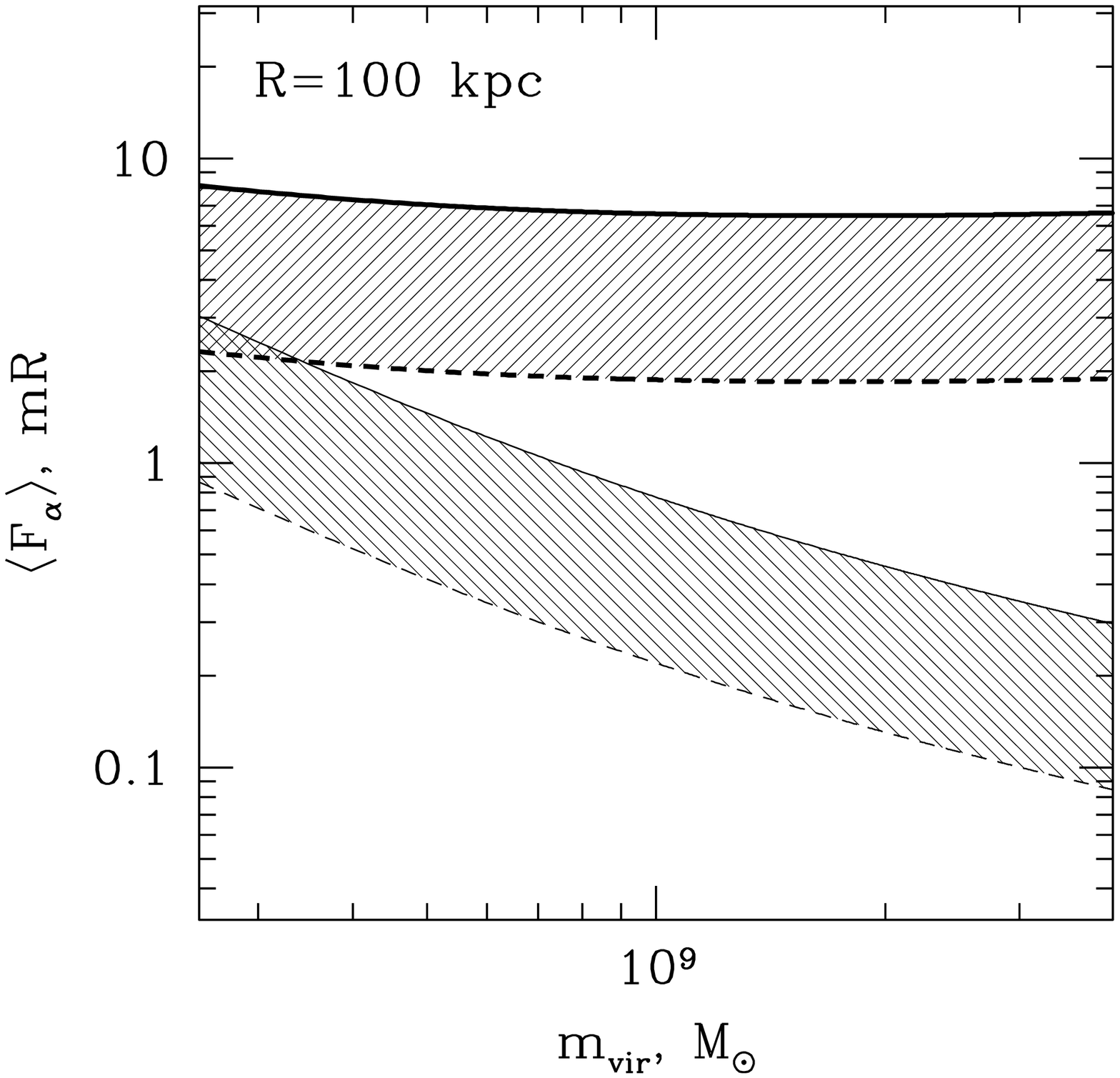}\hspace{30pt}\includegraphics[scale=0.4]{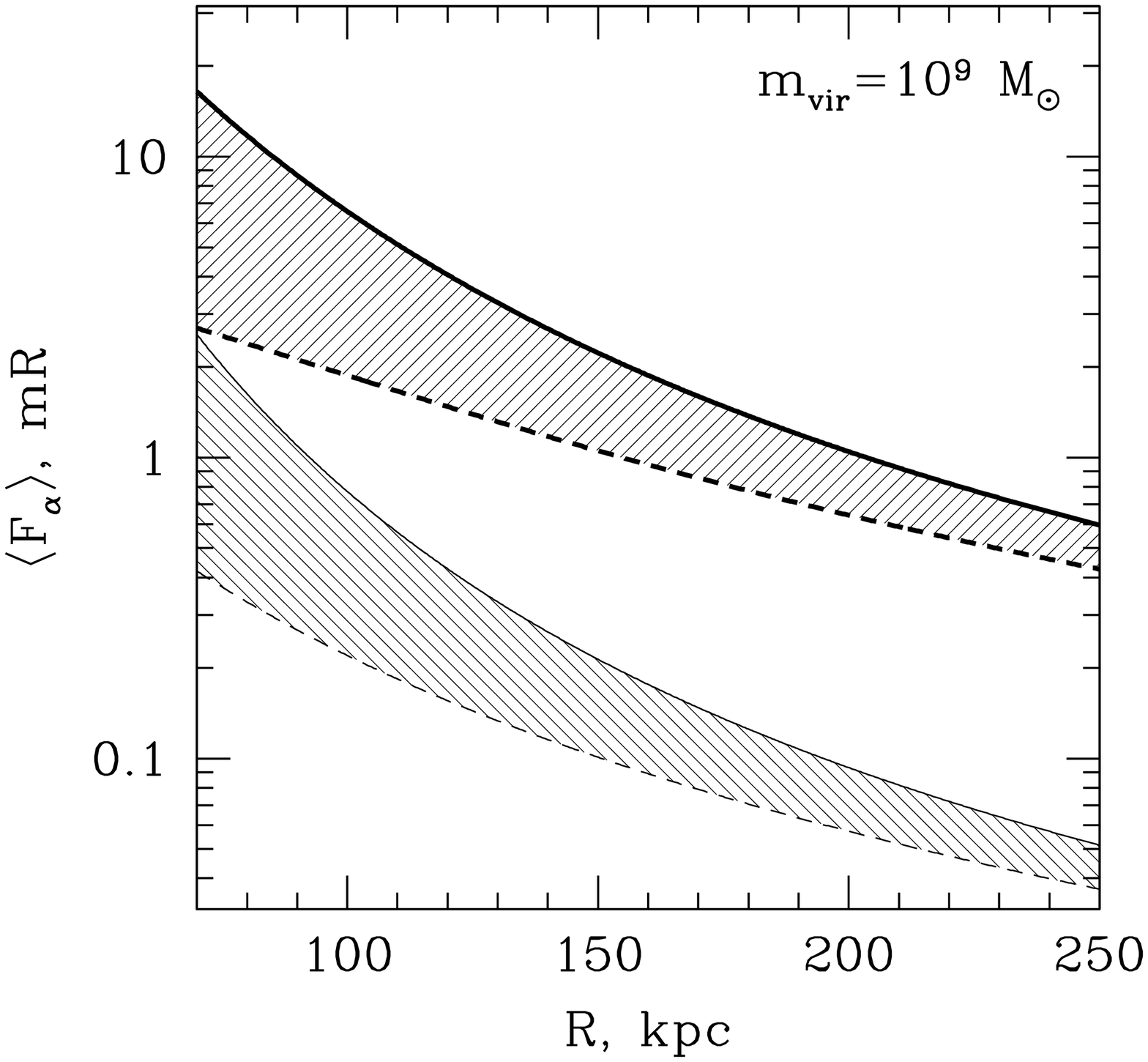}
\caption{
H$_\alpha$ central surface brightness averaged over a $1\degr$ circular beam for
critical ISM in Burkert (thick lines) and NFW (thin lines) halos as a function
of the virial mass (left panel) and galactocentric distance (right panel). Solid
(dashed) lines correspond to the external LyC flux $F_{\rm tot}$ ($F_{\rm bg}$).
The shaded areas show the locus of half-critical halos.
\label{ion_FRH}}
\end{figure*}

To allow a direct comparison of the WHAM observations with our model, in
Fig.~\ref{ion_FRH} we plot the expected H$_\alpha$ flux $\langle F_\alpha
\rangle$ from half-critical ISM of dSphs averaged over a circular beam with
a diameter of $1\degr$ centred on the dwarf. Fig.~\ref{ion_FRH} shows that
the flux is very low, with NFW halos being $\sim 10$ times fainter than Burkert's
halos. For both NFW and Burkert halos with masses $2.5\times 10^8-4\times
10^9$~M$_\odot$, the range of $\langle F_\alpha \rangle$ for half-critical ISM is
$1.3-20.7$~mR for Ursa Minor and $0.6-12.7$~mR for Draco. As one can see, the
whole range of H$_\alpha$ brightness expected from half-critical ISM in Galactic
dSphs is below the best available observational upper limits.

\begin{figure*}
\includegraphics[scale=0.4]{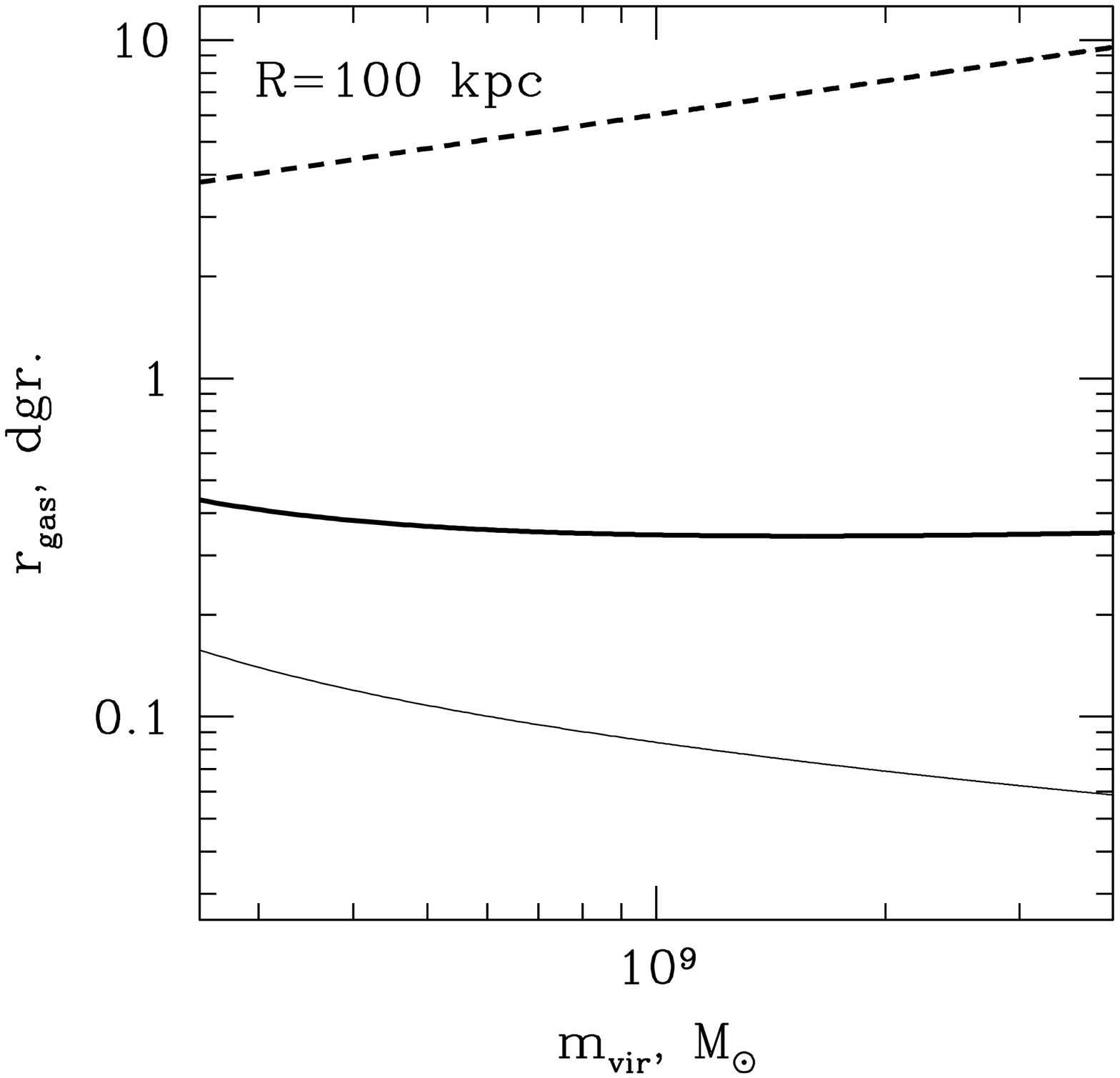}\hspace{30pt}\includegraphics[scale=0.4]{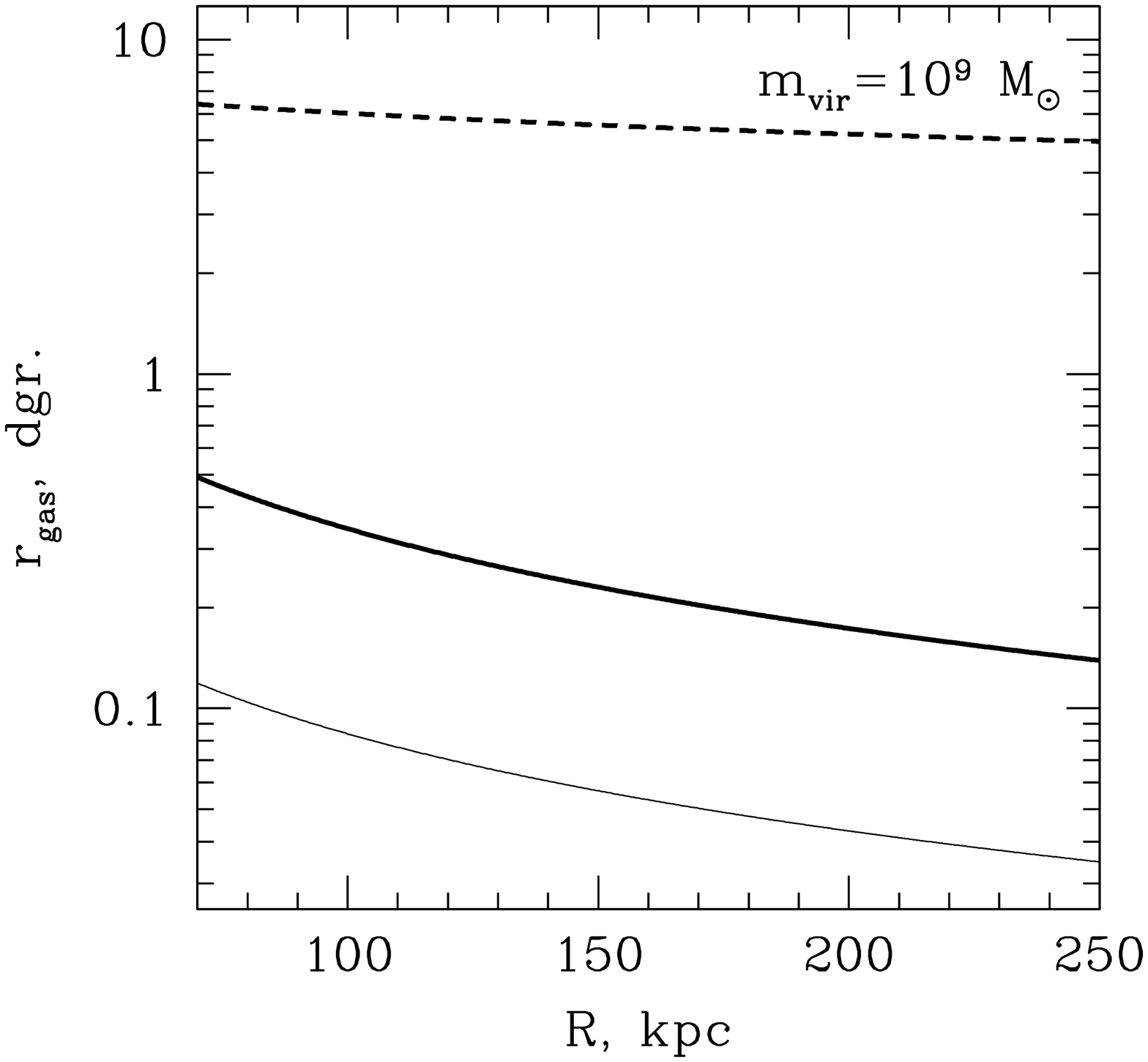}
\caption{
Exponential scalelength in angular units of H$_\alpha$ surface brightness for
fully photoionized ISM in Burkert (thick lines) and NFW (thin lines) halos as a
function of the virial mass (left panel) and galactocentric distance (right
panel). Dashed lines show the apparent tidal radius of DM halos.
\label{ion_Ha2}}
\end{figure*}

Finally, in Fig.~\ref{ion_Ha2} we show the apparent extent of the H$_\alpha$
emission for our model halos. Here we plot the angular distance $r_{\rm gas}$
from the centre of the dwarf where the H$_\alpha$ surface brightness drops by a
factor of $e$. For our range of $m_{\rm vir}$ and $R$, the angular scalelength
of H$_\alpha$ emission ranges between $1\farcm 5$ (NFW halo with $m_{\rm
vir}=4\times 10^9$~M$_\odot$ and $R=250$~kpc) and $37'$ (Burkert halo with
$m_{\rm vir}= 2.5\times 10^8$~M$_\odot$ and $R=70$~kpc).

The conclusion we draw here is that to rule out the presence of fully
photoionized ISM in half-critical state in Galactic dwarf spheroidals, either
much higher sensitivity (down to $0.1-0.2$~mR) low angular resolution ($\sim
1\degr$), or lower sensitivity ($\sim 5$~mR) but much higher angular resolution
($\sim 1'$) H$_\alpha$ observations are required.

\subsection{FUV radiation}
\label{FUV}

FUV radiation between 912 and 2000~\AA, absorbed by interstellar dust, is the
most important source of heating for the Galactic neutral ISM \citep{wol95}.
The UV heating is proportional to the density of the gas. The denser CNM phase
of the ISM is thus more sensitive to changes in the level of the FUV radiation
than the WNM phase, and for sufficiently large FUV fluxes can be completely
evaporated (CNM$\rightarrow$WNM phase transition). The presence of CNM is
believed to be essential for molecular hydrogen cloud formation, and hence for
star formation.

Even the smallest isolated dIrr galaxies are known to possess two-phase ISM
\citep{you96,you97b}. Theoretical models predict that a very low level
radiation field, 0.001--0.01 of the local Galactic value, produced by stars in a
low mass dIrr can maintain its ISM in a two-phase state at pressures $\sim
10-200$~K~cm$^{-3}$ \citep[their fig.~16]{you97b}. If such a dwarf is exposed
to external FUV radiation (e.g. from a nearby spiral galaxy) with the flux larger
than the internal one, the balance between CNM/WNM phases of the ISM will be
shifted toward the WNM phase, reducing the SFR. For large enough external FUV
flux the star formation will be quenched. Thus it appears that the FUV radiation
from giant spirals could be another important environmental factor for
intrinsically faint satellite galaxies.

We consider the two following main components of the FUV radiation field inside
early-type dwarfs: internal (from the Population II stars), and external (from
the host spiral galaxy).

From table~4 of \citet{wel96} we derive the following estimate of the intensity
of the FUV interstellar radiation from the old population stars averaged over
the wavelength interval $900-1100$~\AA~ at the centre of the dwarf elliptical
galaxy NGC~185: $\sim 2000$~cm$^{-2}$~s$^{-1}$~\AA$^{-1}$.  We assume that the
FUV flux at the centre of a dwarf scales as in a homogeneous stellar sphere as
$I_0^{2/3}\,L_V^{1/3}$, where $I_0$ is the central V-band luminosity density and
$L_V$ is the V-band luminosity of the galaxy. For NGC~185,
\citet{mat98} gives the following values: $I_0=1.76$~L$_{\odot}$~pc$^{-3}$ and
$L_V=1.25\times 10^8$~L$_{\odot}$. We obtain the following estimate of the FUV
flux at the centre of early-type dwarfs, expressed in units of the standard
local FUV background with the flux $F_{loc}=1.0\times
10^5$~cm$^{-2}$~s$^{-1}$~\AA$^{-1}$ at $\lambda=1000$~\AA~ (\citealt{dra78},
their equation~[11]):

\begin{equation}
\label{T2}
f_{int}\simeq 2.7\times 10^{-5}\,I_0^{2/3}\,L_V^{1/3},
\end{equation}

\noindent where $I_0$ is in L$_{\odot}$~pc$^{-3}$, and $L_V$ is in L$_{\odot}$.

The FUV luminosity of the Local Group giant spirals (Milky Way and M31) is not
well known. We assume that, for face-on giant spiral galaxies, the FUV flux is
proportional to the H$_{\alpha}$ flux (because most of both types of radiation
can be traced back to the same source --- young massive stars). Two nearby
face-on spiral galaxies, similar to M31 and the Milky Way in their Hubble type
and luminosity, have been observed with the Ultraviolet Imaging Telescope (UIT,
\citealt{ste97}) aboard the space shuttles: NGC~628 (M74) and NGC~5457 (M101).
\citet{bel01}
presented for these galaxies both FUV flux at $\lambda=1521$~\AA~ (UIT filter
B1), and H$_{\alpha}$ flux. In accord with our expectations, the ratio of the
H$_{\alpha}$ flux to the FUV flux is comparable for the two spirals: $f_{{\rm
H}_{\alpha}}/f_{\rm 1521}=14$~\AA~ for NGC~628, and 10~\AA~ for NGC~5457 (the
units for $f_{{\rm H}_{\alpha}}$ are ergs~cm$^{-2}$~s$^{-1}$, and for $f_{\rm
1521}$ are ergs~cm$^{-2}$~s$^{-1}$~\AA$^{-1}$ --- hence the units of
\AA~ for the ratio of the fluxes). The average value of the ratio is

\begin{equation}
\label{T3}
\frac{f_{{\rm H}_{\alpha}}}{f_{\rm 1521}} \simeq 12\, \mbox{\AA}.
\end{equation}

The observed H$_{\alpha}$ flux from M31 is $f_{{\rm H}_{\alpha}}\simeq 5.1\times
10^{-10}$~ergs~cm$^{-2}$~s$^{-1}$ \citep{dev94}.  M31 is a highly inclined
spiral with $i=77\fdg 5$ \citep*{map97}.  Adopting the average ${\rm
H}_{\alpha}$ optical depth value $\tau_{{\rm H}_{\alpha}}=1.23$ of \citet{wal94}
and assuming that the absorbing layer is thin, the deprojected face-on
H$_{\alpha}$ flux from M31 is $\exp [\tau_{{\rm H}_{\alpha}}(1-\cos i)]\simeq
2.6$ times larger than the observed one, or $f'_{{\rm H}_{\alpha}} \simeq
1.3\times 10^{-9}$~ergs~cm$^{-2}$~s$^{-1}$.  Using the empirical conversion
factor (equation~[\ref{T3}]), we estimate the M31 face-on flux in B1 filter as
$f'_{1521}
\simeq 1.1\times 10^{-10}$~ergs~cm$^{-2}$~s$^{-1}$~\AA$^{-1}\simeq
8.3$~cm$^{-2}$~s$^{-1}$~\AA$^{-1}$. We adopt the distance to M31 of 780~kpc (see
discussion in Section~\ref{twins}).  The FUV flux from M31 along its polar axis
at the distance $R_{\rm kpc}$ expressed in units of the standard local FUV
background with the flux $F_{loc}=2.0\times 10^5$~cm$^{-2}$~s$^{-1}$~\AA$^{-1}$
at $\lambda=1521$~\AA~
\citep{dra78} is then:

\begin{equation}
\label{T4}
f_{ext}\simeq 25\,R_{\rm kpc}^{-2}.
\end{equation}

Finally, dividing equation~(\ref{T2}) by equation~(\ref{T4}), we obtain a rough
estimate of the ratio of the internal FUV flux to the flux from the host galaxy
at the centre of the early-type dwarf satellites of M31 and the Milky Way
(assuming, that the Milky Way FUV luminosity is comparable to that of M31):

\begin{equation}
\label{T5}
\frac{f_{int}}{f_{ext}}\simeq 1.1\times 10^{-6}\,I_0^{2/3}L_V^{1/3}R_{\rm kpc}^{2}.
\end{equation}

\noindent This equation is applicable to dwarfs located far from the plane of the host
galaxy, where the dust attenuation becomes significant.

Using equation~(\ref{T5}) and the data on dSphs from the review of
\citet{mat98}, we can estimate the relative importance of the external FUV
radiation for the Milky Way satellites. For all but one dSphs (Sextans, Ursa
Minor, Draco, Carina, Sculptor, Fornax, and Leo~II) the internal FUV flux is
found to be much smaller than the external one even at the centre of the dwarf,
with the $f_{int}/f_{ext}$ values ranging from $\sim 0.01$ (Sextans and Ursa
Minor) to $\sim 0.4$ (Fornax and Leo~II).  Leo~I appears to be the only Galactic
dSph for which the internal FUV flux at the centre of the dwarf dominates over
the external one: $f_{int}/f_{ext}\sim 2$.

Equation~(\ref{T4}) can also be used to determine whether the FUV flux from a large
spiral galaxy is strong enough to affect the multi-phase structure of the ISM
(and hence the star formation) in a nearby dwarf galaxy. Fig.~16 of
\citet{you97b} suggests that in low metallicity dwarf galaxies, FUV radiation,
with the flux as low as 0.001 of the local Galactic value, can regulate the
structure of the ISM.  For M31 and the Milky Way, equation~(\ref{T4}) indicates
that the FUV flux will be larger than the above value out to a radius of
158~kpc, which encompasses most of the dwarf spheroidal satellites.

\begin{figure*}
\includegraphics[scale=0.9]{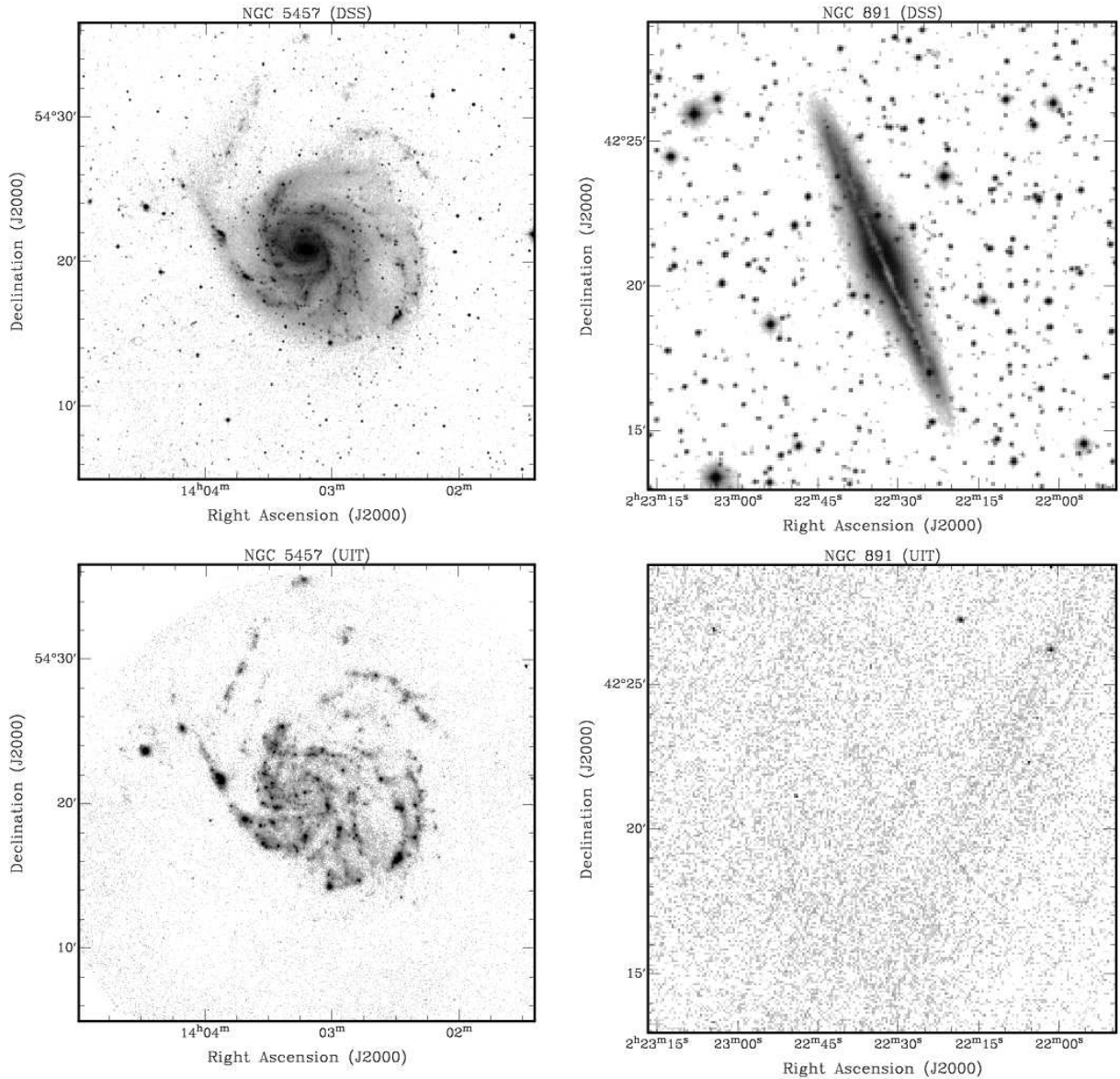}
\caption{Optical DSS (top) and FUV UIT (bottom) images for two M31-like galaxies ---
face-one NGC~5457 (left panels), and edge-on NGC~891 (right panels). UIT images
were obtained with B1 and B5 filters ($\lambda\simeq 1500$~\AA), and have
comparable sensitivity (within a factor of 2).  Weak features in both optical
and FUV images were emphasized by using the square-root intensity scale.  The
fact that NGC~891 is not detected in FUV (bottom right image) points at the strong
anisotropy of the FUV radiation field around normal spiral galaxies.
\label{UIT}}
\end{figure*}

In an M31-like spiral, with low star formation activity and relatively small
bulge, most of the FUV radiation comes from sources in the plane of the
galaxy, closely associated with the spiral arms. The FUV flux from a normal
spiral galaxy is then expected to be anisotropic, with most of the radiation
near the plane of the galaxy being absorbed by the dust. In Fig.~\ref{UIT}, we
use two nearby spirals similar to M31 and the Milky Way, face-on NGC~5457 and
edge-on NGC~891, to illustrate this anisotropy.  A typical dSph galaxy on orbit
around a giant spiral would spend most of its lifetime in the regime
$f_{int}/f_{ext}\ll 1$, with relatively short periods of time of
$f_{int}/f_{ext}\ga 1$ when the dwarf crosses the plane of the host spiral
galaxy every $1-5$~Gyr (half of the orbital period).

The above arguments appear to make a reasonably strong case for the external FUV
radiation as being an important evolutionary factor for low mass dwarfs on
orbits around large spirals.

There are two important caveats to the above analysis. The first one is related to
the poorly known contribution from the metagalactic background to the FUV radiation 
field inside dwarf galaxies. The observed FUV background radiation is known
to be strongly dominated by the local (Galactic) sources, making it almost
impossible to put an accurate upper limit on the metagalactic flux
\citep[see discussion in][]{sas95}. Upper limits for the extragalactic background 
derived by different authors differ by more than an order of magnitude, and are
as low as 380~cm$^{-2}$~s$^{-1}$~\AA$^{-1}$ in $4\pi$ solid angle at
$\lambda=912-1100$~\AA~ \citep{mur99}, or $\sim 0.004$ in the local background
units. Calculations of the propagation of FUV radiation from QSOs and AGNs
through the intergalactic space \citep[their fig.~5a]{haa96} give an estimate
of the lower limit for the background flux, $28$~cm$^{-2}$~s$^{-1}$~\AA$^{-1}$
at $\lambda=1000$~\AA~ ($2.7\times 10^{-4}$ in the local units).  From
equation~(\ref{T4}), this range of possible background values corresponds to
a radius $83<R<310$~kpc for the sphere around M31 where its FUV radiation
dominates over the background (except for the narrow zone near the plane of the
galaxy).

The second caveat is that the heating by soft X-ray radiation, which is inferior
to the FUV heating for the Galactic neutral ISM, could be the dominant heating
mechanism under the low metallicity and low radiation field conditions in dwarf
galaxies \citep{wol95}. We estimate the M31 deprojected face-on X-ray luminosity
in the band $0.1-2$~keV to be $\sim (0.5-1)\times 10^{40}$~ergs~s$^{-1}$. (For
this we used the observed M31 soft X-ray luminosity value of $0.43\times
10^{40}$~ergs~s$^{-1}$ from \citealt{sup01}, recalculated for the distance to
M31 of 780~kpc, and the X-ray luminosity of the face-on spiral NGC~5457
$1.0\times 10^{40}$~ergs~s$^{-1}$ from \citealt{rea01}, recalculated for the
distance of 7.0~Mpc from
\citealt{ste98}.)  The extragalactic soft X-ray background flux is believed to
be known relatively accurately. \citet{che97} give the following estimate of the
background spectral density within the $0.1-7$~keV band:
$10.5\,E^{-1.46}$~keV~cm$^{-2}$~s$^{-1}$~sr$^{-1}$~keV$^{-1}$, where $E$ is the
photon energy in keV units. Integrated over the $0.1-2$~keV interval of photon
energies, this gives the following value for the metagalactic soft X-ray
background flux (in $2\pi$ solid angle): $5.0\times
10^{-7}$~ergs~cm$^{-2}$~s$^{-1}$. Combined with the derived above face-on X-ray
luminosity of M31, the radius of the soft X-ray ``dominance sphere'' for M31 is
found to be very small: $\sim 9-13$~kpc. Thus the ISM heating by soft X-ray
radiation from normal spiral galaxies cannot be an important environmental
factor for dwarf galaxies.

This leads to the conclusion that FUV radiation escaping from spiral galaxies
can be an important environmental factor for dwarf satellite galaxies only if
the resultant ISM heating rate is larger than the heating rate from the
metagalactic soft X-ray background. This issue can be addressed only through
solving numerically the equations of thermal and ionization equilibrium with the
inclusion of all the relevant physical processes (similarly to \citealt{wol95}),
which is beyond the scope of this paper.

\section{OBSERVATIONAL EVIDENCE}
\label{Add}

The previous section gives support to the idea that the electromagnetic
radiation escaping from the host spiral galaxy can be an important evolutionary
factor for dwarf galaxies. This mechanism can explain in general the principal
differences (in SFHs and neutral gas content) between two classes of dwarfs ---
dSphs and dIrrs. In this section we will present observational evidences for
the impact of the UV radiation from the Milky Way and M31 on their dwarf
satellites, which gives further support to the above idea.

\subsection{The puzzle of the ``twin'' galaxies NGC~147 and NGC~185}
\label{twins}

The two dwarf elliptical satellites of M31, NGC~147 and NGC~185, appear to be
almost identical in many respects \citep{you97a}: they have comparable blue
luminosities, Holmberg diameters, {\it B--V} colors, average surface brightness,
light profile shapes, mean stellar dispersions, and projected distances from
M31. Both galaxies show evidence for an intermediate-age stellar component
\citep{you97a}.

Despite all these similarities, NGC~147 and NGC~185 are strikingly different in
their recent SFHs, and in the present day neutral gas content. NGC~185 formed
stars as recently as 20~Myr ago \citep{lee93a}, and contains substantial amounts
of \HI ($1.1\times 10^5$~M$_{\odot}$) and H$_2$ ($4.1\times 10^4$~M$_{\odot}$)
\citep{you01}. Conversely, NGC~147 has not formed stars for at least a Gyr
\citep{han97}, and appears to be devoid of neutral ISM
\citep*{you97a,sag98}.

We argue that the differences between NGC~147 and NGC~185 can be explained by
different fluxes of the LyC and FUV radiation escaping from M31 at the present
locations of the dwarfs.

To quantify this effect, we need to know the accurate distances to the dwarfs
and to M31. To avoid large systematic errors, we use the distance measurements
obtained with the same technique --- the tip of the red giant branch (TRGB)
method, which was shown to work very well for low-metallicity Population II
stars \citep*{lee93b}. Two available TRGB distance measurements for NGC~185
(\citealt{lee93b} and \citealt{mar98}) give virtually the same values of the
true distance modulus: $(m-M)_0=23.95\pm 0.1$. The situation with NGC~147 is
different --- \citet{lee93b} gave the value $(m-M)_0=24.13\pm 0.1$, whereas
\citet{han97} obtained $(m-M)_0=24.37\pm 0.06$. We choose to use the latter value,
because it was derived from the high quality WFCP2 data for $\sim 117,000$ stars
in two different fields, both producing identical true distance modulus
estimates. In contrast, the distance value of \citet{lee93b} was based on the
photometry of only $\sim 500$ stars from the paper of
\citet*{mou83}. Two relevant distance measurements for M31 agree very well:
\citet{dur01} obtained a TRGB true distance modulus $(m-M)_0=24.47\pm 0.12$ based on 
wide-field photometry of a field in the outer halo of M31, and \citet{hol98}
derived $(m-M)_0=24.47\pm 0.07$ by fitting theoretical isochrones to the
observed red giant branches of 14 globular clusters in M31. (It is interesting
to note that an identical value was also obtained by \citet{sta98} from
comparing the red clump stars with parallaxes known to better than 10\% in the
Hipparcos catalog with the red clump stars observed in three fields in M31 using
the Hubble Space Telescope: $(m-M)_0=24.47\pm 0.08$.)  Finally, we adopt the
following values of the true distance modulus for NGC~185, NGC~147, and M31:
$23.95\pm 0.1$, $24.37\pm 0.06$, and $24.47\pm 0.07$, respectively.

We derive the following distances from M31: $100^{+13}_{-4}$~kpc for NGC~147,
and $186^{+34}_{-33}$~kpc for NGC~185. If the FUV and ionizing radiation from
M31 were isotropic, NGC~147 would receive $2.8^{+0.9}_{-0.8}$ times larger flux
than NGC~185. This difference is probably not large enough to explain the
observed differences in the neutral gas content and recent SFH of the dwarfs.

It is expected though that the LyC and FUV fluxes from a giant spiral galaxy are
strongly anisotropic, being at maximum along the polar axis (local galactic
latitude $b_{loc}=\pm 90\degr$), and dropping virtually to zero in the plane of
the galaxy ($b_{loc}=0\degr$). To estimate $b_{loc}$ values for NGC~147 and
NGC~185, we adopt the inclination angle of the M31 disk $i=77\fdg 5$ from
\citet{map97}, and the position angle of the galactic line of nodes ${\rm
PA}=37\fdg 7$ from \citet{dev58}. The uncertainty in both angles is assumed to
be $1\degr$. We use the fact that the northwest side of M31 is the near side
\citep[e.g.][]{wal88}, and arbitrarily choose the local northern pole to be on
the opposite side of the M31 disk.

\begin{figure}
\includegraphics[scale=0.6]{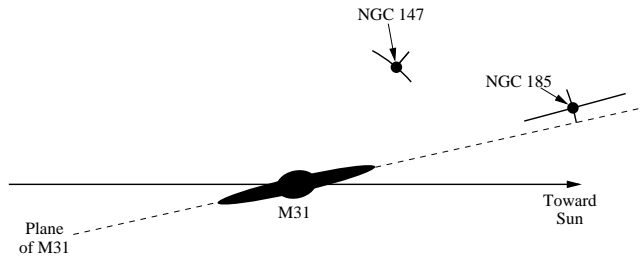}
\caption{
The most probable locations of NGC~147 and NGC~185 relative to the disk of M31
and the observer. The error bars correspond to one-sigma confidence intervals.
Please note that the disk of M31 is not drawn to scale.
\label{fig_N147}}
\end{figure}

Taking into account all the uncertainties, we derive
$b_{loc}=(3^{+4}_{-3})$~dgr.  for NGC~185, and
$b_{loc}=(37^{+11}_{-8})$~dgr. for NGC~147. It appears that NGC~185 is located
almost in the plane of the M31 disk, and is therefore shielded from the FUV and
LyC electromagnetic radiation by the \HI disk of M31. In contrast, from the
location of NGC~147 the M31 spiral is seen half-open (see
Fig.~\ref{fig_N147}), resulting in significant fluxes of the ionizing and FUV
radiation, which can ionize and heat the ISM of the dwarf, thus preventing star
formation.

\subsection{Disturbed ISM of NGC~205}

The neutral ISM of the NGC~205 dwarf elliptical satellite of M31 consists of
$4.3\times 10^5$~M$_{\odot}$ of \HI and $\sim 1\times 10^5$~M$_{\odot}$ of H$_2$
\citep{you00}. The morphology and kinematics of the neutral gas is very unusual:
the central and northern parts of the ISM look relatively unperturbed with the
radial velocity being close to that of the stellar body, whereas the southern
part appears to be compressed in the SE direction and red-shifted by $\sim
20-30$~km~s$^{-1}$ relative to the stars
\citep[their fig.~2 and 4]{you97a}.

\begin{figure}
\hspace{-60pt}\includegraphics[scale=0.6]{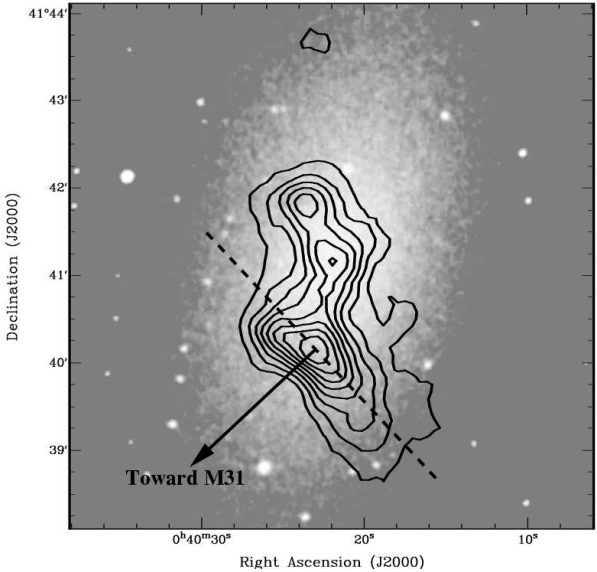}
\caption{
\HI flux contours of NGC~205 from \citet{you97a}, superposed on a DSS image
(only the central brightest part of the optical disk is shown).  The arrow shows
the direction toward the M31 centre at ${\rm PA}=133\degr$. The dashed line is
perpendicular to the arrow. 
\label{fig_N205}}
\end{figure}

\begin{figure}
\includegraphics[scale=0.63]{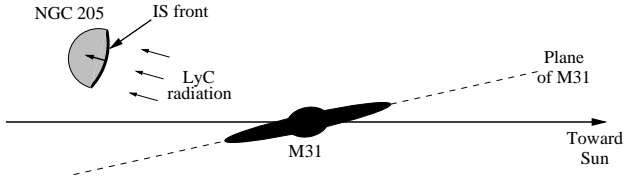}
\caption{
Scheme showing a probable position of NGC~205 relative to M31 and the observer.
In this configuration the IS front in the ISM of NGC~205 would be seen by the
observer red-shifted relative to the stars of the dwarf galaxy.
Please note that the bodies of M31 and NGC~205 are
not drawn to scale.
\label{fig_scheme}}
\end{figure}

We note that the line drawn along the compressed gas in the southern part of the
galaxy is perpendicular to the direction toward the centre of M31, which is
located to the southeast of NGC~205 at the position angle ${\rm PA}=133\degr$
(Fig.~\ref{fig_N205}).  The dwarf galaxy is located very close to M31 in
projection (8~kpc away at the distance of M31), and most probably is $\sim
30-100$~kpc behind the spiral galaxy along the line of sight
\citep{sah92,mat98,bon01}. This places NGC~205 relatively close to the plane of
the spiral galaxy, with the local galactic latitude $b_{loc}\sim 20\degr$. The
morphology and kinematics of the disturbed gas is then consistent with an
ionization-shock (IS) front driven by the LyC radiation from M31 toward the
centre of NGC~205 with a velocity of $\sim 25$~km~s$^{-1}$
(Fig.~\ref{fig_scheme}). The fact that the compressed part of the ISM shows
the highest \HI column density in NGC~205 and contains significant amounts of
H$_2$ kinematically coupled to the \HI \citep{you00} supports this picture of a
relatively strong shock. From Fig.~\ref{fig_N205}, we estimate that the shock
front has already traveled $\sim 300$~pc inside the dwarf, indicating that the
process started very recently: $\sim 10$~Myr ago (kinematic age).

Another important piece of evidence comes from the SFH of the dwarf galaxy.  The
central part of NGC~205 has experienced two recent star bursts --- $20-80$ and
$\sim 500$~Myr ago, with a puzzling gap in between \citep{lee96}.

We propose the following scenario which explains both the ISM and the recent SFH
peculiarities of NGC~205.  We argue that the dwarf galaxy experienced the two
most recent star bursts during its passages through the plane of M31. The star
formation $20-80$~Myr ago, the dwarf's present location close to the plane of
the spiral, and the short time scale $\sim 10$~Myr for the observed IS front all
suggest that NGC~205 has recently crossed the plane of M31. Associating the
second to last star burst $\sim 500$~Myr ago with the previous crossing of the
plane gives an estimate of the orbital period of the dwarf: $P\simeq 1$~Gyr. If
we assume that the dwarf is on circular orbit around M31 with the orbital
velocity $V_c=220$~km~s$^{-1}$, then the radius of the orbit is 36~kpc. At such
a small distance, the LyC and FUV fluxes from M31 should be significant for most
of the lifetime of NGC~205, with only short periods every 500~Myr with much
lower levels of radiation during the passages through the plane of the host
spiral galaxy.

\subsection{Carina dSph versus Ursa Minor dSph}

Carina and Ursa Minor dwarf spheroidals belong to the class of the lowest
luminosity dSph galaxies. They have comparable luminosity ($L_V\sim [3-4]\times
10^5$~L$_{\odot}$), core radius ($r_c\sim 200$~pc), central mass-to-light ratio
($m/L_V \sim 30-60$), and distance from the Milky Way galaxy ($R\sim
70-100$~kpc) \citep{mat98}. It is very surprising then that they exhibit two
opposite SFH extremes: Carina has formed stars for most of its lifetime in three
discrete bursts 3, 7, and 15~Gyr ago \citep{hur98}, whereas Ursa Minor was
formed in a single star burst $\ga 14$~Gyr ago
\citep{ols85,her00}.

We note that Carina has the smallest absolute value of the galactic latitude
$\|b\|\simeq 22\degr$ among all the Milky Way satellite galaxies (excluding the
special case of the Sagittarius dSph). This is consistent with Carina having
relatively small angle between its orbit and the plane of the Galaxy ---
probably as small as $22\degr$. If this is the case, the dwarf has spent a
significant part of its lifetime hiding from the ionizing and FUV radiation from
the Milky Way near the plane of the Galaxy, keeping its ISM neutral and
episodically forming stars.

The opposite example is that of Ursa Minor. Both the proper motion measurements
\citep{sch97} and the relatively high location above the Galactic plane
($\|b\|\simeq 45\degr$) suggest that the dwarf is on a close to polar orbit around
the Milky Way galaxy. Galaxies less massive than $\sim 2\times 10^8$~M$_{\odot}$
should gradually lose their ISM when the gas is fully photoionized. We argue
that the relatively small distance from the Galaxy and the almost polar orbit
caused Ursa Minor to lose its ISM on a short time scale, leading to sharply
declining SFR.

\section{Discussion and conclusions}
\label{Discussion}

Theoretical description of the formation and evolution of dwarf irregular
galaxies is a complex problem, which is far from being fully
resolved. Nevertheless, recent semi-analytical models \citep{spa97,fer00} appear
to converge in that the isolated gas-rich dwarfs should have been forming stars
with almost constant low efficiency for most of their lifetimes, with a possible
exception of the early epoch ($z\ga 1$) with increased SFR.  The emerging
consensus is that soon after the initial star burst, the star formation becomes
self-regulated through the feedback from massive stars (impact on the ISM of
stellar winds, ionizing and FUV radiation, and supernovae). This is in accord
with the findings of \citet{zee01}, who showed that observed optical colors of a
large sample of isolated dIrr galaxies are consistent with approximately
constant SFR for at least 10~Gyr.

The situation with early-type dwarfs (dSph and dE) is even more
complicated. There are at least three observational facts that any successful
dwarf evolution theory should explain:

\begin{enumerate}
\item {\it Impact of environment.} All Local Group dwarfs with $M_{\rm V}>-15^{\rm m}$ located
within 250~kpc from Milky Way and M31 are early-type systems. The observed
morphology-density relation for dwarf galaxies \citep*{bin90} appears to extend
this trend beyond the Local Group.

\item {\it Absence of neutral gas.} With a few exceptions, the Local Group early-type
dwarfs appear to be devoid of \HI and H$_2$ --- despite showing in many cases
the evidence for recent star formation. For example, Fornax has a very low \HI
mass upper limit of $5,000$~M$_{\odot}$
\citep{mat98}, but formed stars as recently as 200~Myr ago \citep*{sav00}.

\item {\it Episodic SFH.} Some Local Group dSph/dE galaxies have puzzling gaps
of $1-5$~Gyr in their SFHs. These time intervals are much longer than the
$10^7-10^8$~yr timescales of internal mechanisms which could effect the star
formation, --- with a possible exception of the SNe~Ia heating mechanism
\citep{bur97}.
\end{enumerate}

In this paper we showed that all the above observational facts can be explained
by the impact of the ultraviolet radiation from the host spiral galaxy on the
ISM of dwarf satellites. We gave evidence in support of the following key
ingredients of our model:

\begin{enumerate}
\item FUV and LyC fluxes from giant spiral galaxies are sufficiently large
to alter the state of the ISM in nearby dwarf galaxies such that star formation
becomes unlikely. The LyC radiation can ionize the ISM through the process of
photoevaporation, making it very hard to detect.

\item The UV radiation from spiral galaxies like the Milky Way and M31 dominates
over the metagalactic background radiation within a large sphere of radius
$\sim 200$~kpc.

\item The ultraviolet radiation field of a spiral galaxy is strongly anisotropic,
with the flux dropping almost to zero in the narrow zone near the plane of the
galaxy.
\end{enumerate}

In our scenario, low mass galaxies orbiting around large spirals spend most of
their lifetimes exposed to the ultraviolet radiation from their host galaxies,
which keeps their ISM in warm ($\sim 10^4$~K) and ionized state.  Twice per
orbital period, each satellite crosses a narrow ``shadow'' zone near the plane
of host galaxy, where the ultraviolet flux is dramatically reduced by the
absorption in the \HI disk of the spiral. During such ``eclipses'', the ISM of
the dwarf galaxy can recombine, cool down, and collapse toward the centre of the
dwarf on timescales of $\sim 100$~Myr. We expect the cold dense gas at the
centre of the dwarf to become self-gravitating, and under favorable conditions
to start forming stars. For a short time, such early-type dwarfs can turn into a
dIrr galaxy, with the two-phase ISM structure, and the low efficiency star
formation self-regulated through the feedback mechanisms. Approaching the edge
of the shadow zone, the dwarf galaxy experiences an increase in the external FUV
flux, which can warm up the ISM and quench the star formation. Finally, the
increased LyC flux can photoionize the warm ISM through the process of
photoevaporation.

An important issue we have to address is whether observations of high-velocity
clouds (HVCs) are consistent with our scenario. HVCs are
\HI clouds which are well separated in radial velocity from the Galactic
\HI disk \citep{wak91}. With a few exceptions, distances to HVCs are either unknown
or strongly model-dependent (like Magellanic Stream). The range of plausible
distances is from a few kpcs to hundreds of kpcs. Many HVCs were detected in
H$_\alpha$ emission line \citep{put03}. Typical H$_\alpha$ emission measures
range from $\sim 30$ to $\sim 600$~mR, which would correspond to the distances
of 14--67~kpc in our simplified description of the Galactic ionizing radiation
field.  Two Sculptor \HI clouds \citep{car98}, that are most probably physically
associated with the Sculptor dSph \citep{bcm03}, would have a low peak
H$_\alpha$ flux of 22~mR in our Galactic LyC model, which is consistent with a
$<26$~mR nondetection of \citet{put03}. Among different hypotheses on the origin
of HVCs, our radiation harassment model is consistent with those that place
these clouds in the Galactic halo at distances $>15$~kpc from the Galactic
centre. The possible scenarios include accretion of warm-hot intergalactic
medium, stripped gas from disrupted Galactic satellites, and gas-rich
DM-dominated subhalos. Recent detection of HVCs around M31 galaxy \citep{thi04}
appears to be consistent with the above scenarios.

One of predictions of our model which can be tested observationally is that at
least some early-type dwarf satellites of the Milky Way and M31 should possess
extended ionized ISM, with the temperature $T\sim 10^4$~K, and low number
density. The total mass of the gas can be comparable to the mass of the
stars. The sound speed in the photoionized gas ($\sim 10$~km~s$^{-1}$) is close
to the stellar velocity dispersion in dSphs, hence both gas and stars are
expected to have a comparable extent (see also Fig.~\ref{ion_Ha2}).  The most
promising candidates are the dwarfs with complex and relatively recent star
formation, located $\la 100$~kpc away from the host galaxy: NGC~147, Fornax, and
Carina.

There are two possible ways to detect such tenuous ionized gas: either in
absorption, by obtaining FUV spectra of QSOs and AGNs located behind the dwarf,
or through its emission (e.g., in H$_\alpha$ line).

The absorption method was used by \citet{bow97}, who put an upper limit on the
column density of the ionized gas around Leo I. They probed with HST three sight
lines toward QSOs passing $\ge 34\arcmin$ from the centre of the dwarf, and did
not detect any absorption in C{\sc~iv}, Si{\sc~ii}, or Si{\sc~iv} spectral lines
at the velocity of Leo I. The photoionized gas in this dwarf (if present) is not
expected to extend far beyond its stellar tidal radius, which is equal to
$12\farcm 6$ \citep{mat98}, so the results of \citet{bow97} cannot be used to
test our model. \citeauthor{bow97} also showed, that with the present day
technology only the brightest quasars, with $m_{\rm V}\la 16^{\rm m}$, can
be used to search for the presence of the tenuous ionized gas in dSphs. To the
best of our knowledge, there are no known QSO/AGN, brighter than $m_{\rm V}\sim
17^{\rm m}$, located in projection within the stellar extent of the Local Group
early-type dwarfs
\citep{tin97,tin99,sch02}.

We believe that the ionized ISM detection in H$_\alpha$ emission line is a more
feasible approach. The expected H$_\alpha$ surface brightness is $<100$~mR, so
the required sensitivity of the observations should be at least 10~mR. The
angular resolution should be at least $\sim 10\arcmin$ for the Milky Way
satellites, and $\sim 1\arcmin$ for the M31 companions. The Galactic foreground
radiation is expected to dominate over the radiation coming from the dwarfs
(with the typical integral flux values of $\sim 1$~R for the high Galactic
latitude dwarfs, and reaching $\sim 20$~R for Carina), so the observations
should be carried out within a narrow, $20-40$~km~s$^{-1}$, interval of the
radial velocities centred at the optical velocity of the dwarf.  Three
available wide-field H$_\alpha$ surveys do not meet the above requirements: two
complementing surveys,
VTSS\footnote{\tt http://www.phys.vt.edu/$\sim$halpha} (Virginia Tech
Spectral-Line Survey, covers northern hemisphere,
\citealt{den98}), and
SHASSA\footnote{\tt http://amundsen.swarthmore.edu/SHASSA} (Southern H-Alpha
Sky Survey Atlas, \citealt{gau01}) lack sensitivity (500~mR) and velocity
information, whereas WHAM\footnote{\tt http://www.astro.wisc.edu/wham} (covers
northern hemisphere, \citealt{rey98}) having reasonably good sensitivity (50~mR)
and spectral resolution ($\sim 10$~km~s$^{-1}$), does not qualify because of the
insufficient angular resolution ($1\degr$) and too narrow velocity coverage:
$V_{\rm LSR}=-100\dots 100$~km~s$^{-1}$. Targeted H$_\alpha$ emission
observations can do a much better job: observations with the sensitivity $\sim
10$~mR, angular and velocity resolutions $\sim 5\arcmin$ and $\sim
10$~km~s$^{-1}$, are almost routine nowadays \citep{wvw01,wey01}. With new
differential techniques, even lower emission levels of $1-2$~mR might be reached
\citep{bla01}.

It is interesting to note, that at least one of the dwarf galaxies in question,
NGC~205, has a tentative extended H$_\alpha$ emission detection in
VTSS\footnote{The Virginia Tech Spectral-Line Survey (VTSS) is supported by the
National Science Foundation.} (field And07). The central part of the dwarf has a
surface brightness of $\sim 5$~R, and the fainter emission with the brightness
of $1-2$~R appears to fill out all the stellar extent of the galaxy. These large
values of the H$_\alpha$ flux, if confirmed, might require alternative
explanations as to the source of ionization.  Future targeted H$_\alpha$
emission observations of NGC~205 are required to resolve the issue.

\section*{Acknowledgments}

We are grateful to Frank Bertoldi, Andrea Ferrara, and Sergey Silich for
insightful discussions and numerous helpful suggestions. We also thank Lisa
Young for providing the \HI data for NGC~205.  C.~C. acknowledges support from
NSERC (Canada) and FCAR (Qu\'ebec).


\begin{thebibliography}{}

\bibitem[\protect\citeauthoryear{Barnes \& de Blok}{2001}]{bar01} Barnes D.~G., de Blok W.~J.~G., 2001, AJ, 122, 825
\bibitem[\protect\citeauthoryear{Bell \& Kennicutt}{2001}]{bel01} Bell E.~F., Kennicutt R.~C., 2001, ApJ, 548, 681
\bibitem[\protect\citeauthoryear{Bertoldi}{1989}]{ber89} Bertoldi F., 1989, ApJ, 346, 735
\bibitem[\protect\citeauthoryear{Bertoldi \& McKee}{1990}]{ber90} Bertoldi F., McKee C.~F., 1990, ApJ, 354, 529
\bibitem[\protect\citeauthoryear{Bianchi, Cristiani \& Kim}{Bianchi et al.}{2001}]{bia01} Bianchi S., Cristiani S.,  Kim T.-S., 2001, A\&A, 376, 1 
\bibitem[\protect\citeauthoryear{Binggeli, Tarenghi \& Sandage}{Binggeli et al.}{1990}]{bin90} Binggeli B., Tarenghi M.,  Sandage A., 1990, A\&A, 228, 42
\bibitem[\protect\citeauthoryear{Binney \& Tremaine}{1994}]{bin94} Binney J., Tremaine S., 1994, Galactic Dynamics (3d ed.). Princeton Univ. Press, Princeton, NJ
\bibitem[\protect\citeauthoryear{Bland-Hawthorn \& Maloney}{2001}]{bla01} Bland-Hawthorn J., Maloney P. R., 2001, in Mulchaey J. S., Stocke J., eds, ASP Conf.~Ser.~Vol.~254, Extragalactic Gas at Low Redshift.  Astron. Soc. Pac., San Fransisco, p. 267
\bibitem[\protect\citeauthoryear{Bond \& Alves}{2001}]{bon01} Bond H.~E., Alves D.~R., 2001, in Szczerba R., Gorny S. K., eds, Astrophysics and Space Science Library Vol. 265, Post-AGB Objects as a Phase of Stellar Evolution. Kluwer Academic Publishers, Boston/Dordrecht/London, p. 77
\bibitem[\protect\citeauthoryear{Bouchard, Carignan \& Mashchenko}{Bouchard et al.}{2003}]{bcm03} Bouchard A., Carignan C., Mashchenko S., 2003, AJ, 126, 1295 
\bibitem[\protect\citeauthoryear{Bowen et al.}{1997}]{bow97} Bowen D.~V., Tolstoy E., Ferrara A., Blades J.~C., Brinks E., 1997, ApJ, 478, 530
\bibitem[\protect\citeauthoryear{Burkert}{1995}]{bur95} Burkert A., 1995, ApJ, 447, L25 
\bibitem[\protect\citeauthoryear{Burkert \& Ruiz-Lapuente}{1997}]{bur97} Burkert A., Ruiz-Lapuente P., 1997, ApJ, 480, 297
\bibitem[\protect\citeauthoryear{Carignan et al.}{1998}]{car98} Carignan C., Beaulieu S., C{\^ o}t{\' e} S., Demers S., Mateo M., 1998, AJ, 116, 1690
\bibitem[\protect\citeauthoryear{Chen, Fabian \& Gendreau}{1997}]{che97} Chen L.-W., Fabian A.~C., Gendreau K.~C., 1997, MNRAS, 285, 449
\bibitem[\protect\citeauthoryear{de Vaucouleurs}{1958}]{dev58} de Vaucouleurs G., 1958, ApJ, 128, 465
\bibitem[\protect\citeauthoryear{Dennison, Simonetti \& Topasna}{1998}]{den98} Dennison B., Simonetti J.~H., Topasna G.~A., 1998, Publications of the Astronomical Society of Australia, 15, 147
\bibitem[\protect\citeauthoryear{Devereux et al.}{1994}]{dev94} Devereux N.~A., Price R., Wells L.~A., Duric N., 1994, AJ, 108, 1667
\bibitem[\protect\citeauthoryear{Draine}{1978}]{dra78} Draine B.~T., 1978, ApJS, 36, 595
\bibitem[\protect\citeauthoryear{Durrell, Harris \& Pritchet}{2001}]{dur01} Durrell P.~R., Harris W.~E., Pritchet C.~J., 2001, AJ, 121, 2557
\bibitem[\protect\citeauthoryear{Ferrara \& Tolstoy}{2000}]{fer00} Ferrara A., Tolstoy E., 2000, MNRAS, 313, 291
\bibitem[\protect\citeauthoryear{Gallagher et al.}{2003}]{gal03} Gallagher J.~S., Madsen G.~J., Reynolds R.~J., Grebel E.~K., Smecker-Hane  T.~A., 2003, ApJ, 588, 326
\bibitem[\protect\citeauthoryear{Gallart et al.}{2001}]{gal01} Gallart C., Mart{\' i}nez-Delgado D., G{\' o}mez-Flechoso M.~A., Mateo M., 2001, AJ, 121, 2572
\bibitem[\protect\citeauthoryear{Gaustad et al.}{2001}]{gau01} Gaustad J.~E., McCullough P.~R., Rosing W., Van Buren D., 2001, PASP, 113, 1326
\bibitem[\protect\citeauthoryear{Grebel}{1997}]{gre97} Grebel E.~K., 1997, Reviews of Modern Astronomy, 10, 29
\bibitem[\protect\citeauthoryear{Haardt \& Madau}{1996}]{haa96} Haardt F., Madau, P., 1996, ApJ, 461, 20
\bibitem[\protect\citeauthoryear{Han et al.}{1997}]{han97} Han M., Hoessel J.~G., Gallagher J.~S., Holtsman J., Stetson P.~B., 1997, AJ, 113, 1001
\bibitem[\protect\citeauthoryear{Hayashi et al.}{2003}]{hay03} Hayashi E., Navarro J.~F., Taylor J.~E., Stadel J., Quinn T., 2003, ApJ, 584, 541 
\bibitem[\protect\citeauthoryear{Hernandez, Gilmore \& Valls-Gabaud}{2000}]{her00} Hernandez X., Gilmore G., Valls-Gabaud D., 2000, MNRAS, 317, 831
\bibitem[\protect\citeauthoryear{Holland}{1998}]{hol98} Holland S., 1998, AJ, 115, 1916
\bibitem[\protect\citeauthoryear{Hurley-Keller, Mateo \& Nemec}{Hurley-Keller et al.}{1998}]{hur98} Hurley-Keller D., Mateo M., Nemec J., 1998, AJ, 115, 1840
\bibitem[\protect\citeauthoryear{Johnston}{1998}]{joh98} Johnston K.~V., 1998, ApJ, 495, 297 
\bibitem[\protect\citeauthoryear{Lee, Freedman \& Madore}{Lee et al.}{1993a}]{lee93a} Lee M.~G., Freedman W.~L., Madore B.~F., 1993a, AJ, 106, 964
\bibitem[\protect\citeauthoryear{Lee, Freedman \& Madore}{Lee et al.}{1993b}]{lee93b} Lee M.~G., Freedman W.~L., Madore B.~F., 1993b, ApJ, 417, 553
\bibitem[\protect\citeauthoryear{Lee}{1996}]{lee96} Lee M.~G., 1996, AJ, 112, 1438
\bibitem[\protect\citeauthoryear{Lefloch \& Lazareff}{1994}]{lef94} Lefloch B., Lazareff B., 1994, A\&A, 289, 559 
\bibitem[\protect\citeauthoryear{{\L}okas}{2002}]{lok02} {\L}okas E.~L., 2002, MNRAS, 333, 697 
\bibitem[\protect\citeauthoryear{Ma, Peng \& Gu}{Ma et al.}{1997}]{map97} Ma J., Peng Q., Gu Q., 1997, ApJ, 490, L51
\bibitem[\protect\citeauthoryear{Mart{\' i}nez-Delgado \& Aparicio}{1998}]{mar98} Mart{\' i}nez-Delgado D., Aparicio A., 1998, AJ, 115, 1462
\bibitem[\protect\citeauthoryear{Mart{\' i}nez-Delgado, Gallart \& Aparicio}{1999}]{mga99} Mart{\' i}nez-Delgado D., Gallart C., Aparicio A., 1999, AJ, 118, 862
\bibitem[\protect\citeauthoryear{Mateo}{1998}]{mat98} Mateo M.~L., 1998, ARA\&A, 36, 435
\bibitem[\protect\citeauthoryear{Mayer et al.}{2001}]{may01} Mayer L., Governato F., Colpi M., Moore B., Quinn T., Wadsley J., Stadel J., Lake G., 2001, ApJ, 559, 754
\bibitem[\protect\citeauthoryear{Miller et al.}{2001}]{mil01} Miller B.~W., Dolphin A.~E., Lee M.~G., Kim S.~C., Hodge P., 2001, ApJ, 562, 713
\bibitem[\protect\citeauthoryear{Mould, Kristian \& Da Costa}{Mould et al.}{1983}]{mou83} Mould J.~R., Kristian J.,  Da Costa G.~S., 1983, ApJ, 270, 471
\bibitem[\protect\citeauthoryear{Murthy et al.}{1999}]{mur99} Murthy J., Hall D., Earl M., Henry R.~C.,  Holberg J.~B., 1999, ApJ, 522, 904
\bibitem[\protect\citeauthoryear{Odenkirchen et al.}{2001}]{ode01} Odenkirchen M.~et al., 2001, AJ, 122, 2538
\bibitem[\protect\citeauthoryear{Olszewski \& Aaronson}{1985}]{ols85} Olszewski E.~W., Aaronson, M., 1985, AJ, 90, 2221
\bibitem[\protect\citeauthoryear{Piatek et al.}{2002}]{pia02} Piatek S.~et al., 2002, AJ, 124, 3198 
\bibitem[\protect\citeauthoryear{Piersimoni et al.}{1999}]{pie99} Piersimoni A.~M.,  Bono G., Castellani M., Marconi G., Cassisi S., Buonanno R., Nonino M., 1999, A\&A, 352, L63
\bibitem[\protect\citeauthoryear{Putman et al.}{2003}]{put03} Putman M.~E., Bland-Hawthorn J., Veilleux S., Gibson B.~K., Freeman K.~C., Maloney P.~R., 2003, ApJ, 597, 948 
\bibitem[\protect\citeauthoryear{Read \& Ponman}{2001}]{rea01} Read A.~M., Ponman, T.~J., 2001, MNRAS, 328, 127
\bibitem[\protect\citeauthoryear{Reynolds et al.}{1998}]{rey98} Reynolds R.~J., Tufte S.~L., Haffner L.~M., Jaehnig K.,  Percival J.~W., 1998, Publications of the Astronomical Society of Australia, 15, 14
\bibitem[\protect\citeauthoryear{Sage, Welch \& Mitchell}{Sage et al.}{1998}]{sag98} Sage L.~J., Welch G.~A.,  Mitchell G.~F., 1998, ApJ, 507, 726
\bibitem[\protect\citeauthoryear{Saha, Hoessel \& Krist}{Saha et al.}{1992}]{sah92} Saha A., Hoessel J.~G.,  Krist J., 1992, AJ, 103, 84
\bibitem[\protect\citeauthoryear{Sasseen et al.}{1995}]{sas95} Sasseen, T.~P., Lampton, M., Bowyer, S.,  Wu, X., 1995, ApJ, 447, 630
\bibitem[\protect\citeauthoryear{Saviane, Held \& Bertelli}{Saviane et al.}{2000}]{sav00} Saviane I., Held E.~V.,  Bertelli G., 2000, A\&A, 355, 56
\bibitem[\protect\citeauthoryear{Schneider et al.}{2002}]{sch02} Schneider D.~P.~et al., 2002, AJ, 123, 567
\bibitem[\protect\citeauthoryear{Schweitzer, Cudworth \& Majewski}{Schweitzer et al.}{1997}]{sch97} Schweitzer A.~E., Cudworth K.~M.,  Majewski S.~R., 1997, in Humphreys  R. M., ed., ASP Conf.~Ser.~Vol.~127, Proper Motions and Galactic Astronomy. Astron. Soc. Pac., San Fransisco, p. 103
\bibitem[\protect\citeauthoryear{Shull et al.}{1999}]{shu99} Shull J.~M., Roberts D., Giroux M.~L., Penton S.~V.,  Fardal M.~A., 1999, AJ, 118, 1450
\bibitem[\protect\citeauthoryear{Silk, Wyse \& Shields}{Silk et al.}{1987}]{sil87} Silk J., Wyse R.~F.~G.,  Shields G.~A., 1987, ApJ, 322, L59
\bibitem[\protect\citeauthoryear{Spaans \& Norman}{1997}]{spa97} Spaans M., Norman, C.~A., 1997, ApJ, 483, 87
\bibitem[\protect\citeauthoryear{Spitzer}{1978}]{spi78} Spitzer L., Jr., 1978, Physical Processes in the Interstellar Medium. Wiley-Interscience, New York
\bibitem[\protect\citeauthoryear{St-Germain et al.}{1999}]{stg99} St-Germain J., Carignan C., C{\^ o}te S.,  Oosterloo T., 1999, AJ, 118, 1235
\bibitem[\protect\citeauthoryear{Stanek \& Garnavich}{1998}]{sta98} Stanek K.~Z., Garnavich P.~M., 1998, ApJ, 503, L131
\bibitem[\protect\citeauthoryear{Stecher et al.}{1997}]{ste97} Stecher T.~P.~et al., 1997, PASP, 109, 584
\bibitem[\protect\citeauthoryear{Sternberg, McKee \& Wolfire}{Sternberg et al.}{2002}]{ste02} Sternberg A., McKee C.~F.,  Wolfire M.~G., 2002, ApJS, 143, 419 
\bibitem[\protect\citeauthoryear{Stetson et al.}{1998}]{ste98} Stetson P.~B.~et al., 1998, ApJ, 508, 491
\bibitem[\protect\citeauthoryear{Stoehr et al.}{2002}]{sto02} Stoehr F., White S.~D.~M., Tormen G.,  Springel V., 2002, MNRAS, 335, L84
\bibitem[\protect\citeauthoryear{Supper et al.}{2001}]{sup01} Supper R., Hasinger G., Lewin W.~H.~G., Magnier E.~A., van Paradijs J., Pietsch W., Read A.~M.,  Tr{\" u}mper J., 2001, A\&A, 373, 63
\bibitem[\protect\citeauthoryear{Sutherland \& Dopita}{1993}]{sut93} Sutherland R.~S., Dopita M.~A., 1993, ApJS, 88, 253 
\bibitem[\protect\citeauthoryear{Thilker et al.}{2004}]{thi04} Thilker D.~A., Braun R., Walterbos R.~A.~M., Corbelli E., Lockman F.~J., Murphy E., Maddalena R., 2004, ApJ, 601, L39 
\bibitem[\protect\citeauthoryear{Tinney, Da Costa \& Zinnecker}{1997}]{tin97} Tinney C.~G., Da Costa G.~S.,  Zinnecker H., 1997, MNRAS, 285, 111
\bibitem[\protect\citeauthoryear{Tinney}{1999}]{tin99} Tinney C.~G., 1999, MNRAS, 303, 565
\bibitem[\protect\citeauthoryear{van den Bergh}{1999}]{ber99} van den Bergh S., 1999, AJ, 117, 2211
\bibitem[\protect\citeauthoryear{van Zee}{2001}]{zee01} van Zee L., 2001, AJ, 121, 2003
\bibitem[\protect\citeauthoryear{Wakker \& van Woerden}{1991}]{wak91} Wakker B.~P., van Woerden H., 1991, A\&A, 250, 509 
\bibitem[\protect\citeauthoryear{Walterbos \& Braun}{1994}]{wal94} Walterbos R.~A.~M., Braun R., 1994, ApJ, 431, 156
\bibitem[\protect\citeauthoryear{Walterbos \& Kennicutt}{1988}]{wal88} Walterbos R.~A.~M., Kennicutt R.~C., 1988, A\&A, 198, 61
\bibitem[\protect\citeauthoryear{Weiner, Vogel \& Williams}{Weiner et al.}{2002}]{wvw01} Weiner B.~J., Vogel S.~N.,  Williams T.~B., 2002, in  Mulchaey J. S., Stocke J., eds, ASP Conf.~Ser.~Vol.~254, Extragalactic Gas at Low Redshift. Astron. Soc. Pac., San Fransisco, p. 256
\bibitem[\protect\citeauthoryear{Welch, Mitchell \& Yi}{1996}]{wel96} Welch G.~A., Mitchell G.~F.,  Yi S., 1996, ApJ, 470, 781
\bibitem[\protect\citeauthoryear{Weymann et al.}{2001}]{wey01} Weymann R.~J., Vogel S.~N., Veilleux S.,  Epps H.~W., 2001, ApJ, 561, 559
\bibitem[\protect\citeauthoryear{Wolfire et al.}{1995}]{wol95} Wolfire M.~G., Hollenbach D., McKee C.~F., Tielens A.~G.~G.~M.,  Bakes E.~L.~O., 1995, ApJ, 443, 152
\bibitem[\protect\citeauthoryear{Young \& Lo}{1996}]{you96} Young L.~M., Lo K.~Y., 1996, ApJ, 462, 203
\bibitem[\protect\citeauthoryear{Young \& Lo}{1997a}]{you97a} Young L.~M., Lo K.~Y., 1997a, ApJ, 476, 127
\bibitem[\protect\citeauthoryear{Young \& Lo}{1997b}]{you97b} Young L.~M., Lo K.~Y., 1997b, ApJ, 490, 710
\bibitem[\protect\citeauthoryear{Young}{2000}]{you00} Young L.~M., 2000, AJ, 120, 2460
\bibitem[\protect\citeauthoryear{Young}{2001}]{you01} Young L.~M., 2001, AJ, 122, 1747
\end{thebibliography}
\end{document}